\documentclass[preprint,showpacs,showkeys,preprintnumbers,amsmath,amssymb,prb,superscriptaddress]{revtex4} 
\usepackage{epsfig} 
\usepackage{graphicx}
\usepackage{dcolumn}
\usepackage{bm}
\usepackage[usenames]{color}
\usepackage{ulem} 
\usepackage{amsmath}
\usepackage{amssymb}

\begin{document}


\title{Accurate Noise Projection for Reduced \\Stochastic Epidemic Models}

\author{Eric Forgoston\footnote{Corresponding author:  eric.forgoston.ctr@nrl.navy.mil}}
\affiliation{Nonlinear Dynamical Systems Section, Plasma Physics Division,
  Code 6792,\\ U.S. Naval Research Laboratory, Washington, DC 20375, USA}

\author{Lora Billings}
\affiliation{Department of Mathematical Sciences, Montclair State University,
  1 Normal  Avenue, Montclair, NJ 07043, USA}

\author{Ira B. Schwartz}
\affiliation{Nonlinear Dynamical Systems Section, Plasma Physics Division,
  Code 6792,\\ U.S. Naval Research Laboratory, Washington, DC 20375, USA}



\begin{abstract}
We consider a stochastic Susceptible-Exposed-Infected-Recovered (SEIR)
epidemiological model.  Through the use of a normal form coordinate transform,
we are able to analytically derive the stochastic center manifold along with
the associated, reduced set of stochastic evolution equations.  The
transformation correctly projects both the dynamics and  the noise onto the center manifold.  Therefore, the solution
of this reduced stochastic dynamical system yields excellent agreement, both
in amplitude and phase, with
the solution of the original stochastic system for a temporal scale that is orders of
magnitude longer than the typical relaxation time.  This new method allows for
improved time series prediction of the number of infectious
cases when modeling the spread of disease
in a population.  Numerical solutions of the fluctuations of the SEIR model
are considered in the infinite population limit using a Langevin equation
approach, as well as in a finite population simulated as a Markov process.    
\end{abstract}



\maketitle

{\bf The reduction of high-dimensional, stochastic systems is an important
  and fundamental problem in nonlinear dynamical systems.  In this article, we
  present a general theory of stochastic model reduction which is based on a
  normal form coordinate transform.  This nonlinear, stochastic projection allows for the deterministic and stochastic dynamics to interact correctly on the lower-dimensional manifold so that the dynamics predicted by the reduced,
stochastic system agree well with the dynamics predicted by the original,
high-dimensional, stochastic system.  Although the method may be applied to
any physical or biological system with well-separated time scales, here we
apply the method to an epidemiological model.  We show that when compared to
the original, stochastic epidemic model, the reduced model properly captures
the initial and recurrent disease outbreaks, both in amplitude and phase.  This
long-term accuracy of the reduced model allows for the application of
effective disease control where phase differences between outbreak times and
vaccine controls are important.  Additionally, in practice, one can only
measure the number of infectious individuals in a population.  Our method
allows one to predict the number of unobserved exposed individuals based on
the observed number of infectious individuals.}

\section{Introduction}

The interaction between deterministic and stochastic effects in population dynamics has
played, and continues to play, an important role in the modeling of
infectious diseases. The mechanistic modeling side of population dynamics
is well-known and established~\cite{Anderson91,bailey75}.  These
models typically are assumed to be useful for infinitely large, homogeneous
populations, and arise from the mean field analysis of probabilistic
models. On the other hand, when one considers finite populations, random interactions
give rise to internal noise effects, which may introduce new dynamics.  Stochastic
effects are quite prominent in finite populations, which can range from
ecological dynamics~\cite{Marion2000} to childhood epidemics in
cities~\cite{Nguyen2008,Rohani2002}.  For homogeneous populations
with seasonal forcing, noise also comes into play in the prediction
of large outbreaks~\cite{Rand1991,BillingsBS02,stolhu07}.  Specifically, external random perturbations change the probabilistic
prediction of epidemic outbreaks as well as its control~\cite{Schwartz2004}. 

When geometric structure is applied to the population, the interactions
are modeled as a network~\cite{Pastor-SatorrasV01,moreno2002}.  Many
types of static networks which support epidemics have been considered.  Some
examples include small world networks~\cite{Vazquez06}, hierarchical networks~\cite{WattsMMD05}, and transportation networks of patch
models~\cite{ColizzaBBV06}. In addition, the fluctuation of epidemics
on adaptive networks, where the wiring between nodes changes in response to the
node information, has been examined~\cite{shasch08}.  In adaptive
network models, even the mean field can be
high-dimensional, since nodes and links evolve in time and must be approximated as an additional set of ordinary differential equations. 

Another aspect of epidemic models which is often of interest involves the
inclusion of a time delay.  The delay term makes the analysis significantly
more complicated.  However, it is possible to approximate the delay by
creating a cascade consisting of a large number of compartments~\cite{mrv05}.  For example,
one could simulate the delay associated with a disease exposure time with
several hundred ``exposed'' compartments. 

These model examples are just a few of the types of very high-dimensional
models that are currently of interest.  As a result of the high-dimensionality,
there is much computation involved, and the analysis is quite difficult.  In
particular, real-time computation is not currently possible.  However,
there are usually many time scales that are well-separated (due typically to a
large range in order of magnitude of the parameters) when 
considering such high-dimensional problems.  In the presence of well-separated time scales, a model reduction method is needed to examine
the dynamics on a lower-dimensional space.  It is known that deterministic
model reduction methods may not work well in the stochastic realm, which
includes epidemic models~\cite{forsch09}.  The purpose of this article is to examine
a method of nonlinear, stochastic projection so that the deterministic
and stochastic dynamics interact correctly on the lower-dimensional
manifold and predict correctly the dynamics when compared to the full
system. Because the noise affects the timing of outbreaks, it is essential
to produce a low-dimensional system which captures the correct timing
of the outbreaks as well as the amplitude and phase of any recurrent behavior.

We will demonstrate that our stochastic model reduction method properly
captures the initial disease outbreak and continues to accurately predict the
outbreaks for time scales which are orders of magnitude longer than the
typical relaxation time.  Furthermore, in practice, real disease data includes
only the number of infectious individuals.  Our method allows us to predict the number of
unobserved exposed individuals based on the observed number of infectious individuals.

For stochastic model reduction, there exist several potential methods
for general problems.  For a system with certain spectral requirements,
the existence of a stochastic center manifold was proven in Ref.~\onlinecite{box89}.  Non-rigorous stochastic normal form analysis (which leads to the stochastic
center manifold) was performed in Refs.~\onlinecite{coelti85,knowie83,nam90,namlin91}.  Rigorous theoretical analysis of normal form coordinate transformations
for stochastic center manifold reduction was developed
in Refs.~\onlinecite{arn98,arnimk98}.  Later, another method of stochastic normal form reduction was developed~\cite{rob08},
in which any anticipatory convolutions (integrals into the future
of the noise processes) that appeared in the slow modes were removed.  Since this latter analysis makes the construction of the stochastic
normal form coordinate transform more transparent, we use this method
to derive the reduced stochastic center manifold equation.

\begin{figure}[h!]
\begin{center}
\includegraphics[width=8.5cm,height=4.25cm]{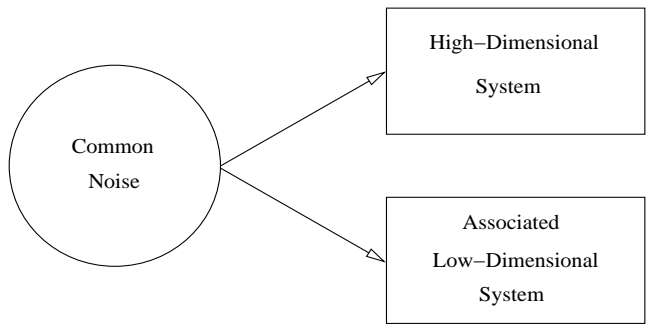}
\caption{\label{fig:noise_cartoon}Schematic demonstrating the injection of a
  common noise into both the high-dimensional system and its associated
  low-dimensional system.}
\end{center}
\end{figure}

Figure~\ref{fig:noise_cartoon} shows a schematic demonstrating our approach to
the problem.  We consider a high-dimensional system along with its
corresponding reduced low-dimensional system.  In this article, two types of
low-dimensional system are discussed:  a reduced system based on deterministic
center manifold analysis and a reduced system based on a stochastic normal
form coordinate transform.  Regardless of the type of low-dimensional system
being considered, a common noise is injected into both the high-dimensional
and low-dimensional models, and an analysis of the solutions found using the
high and low-dimensional systems is performed.

In this article, as a first study of a high-dimensional system, we consider the Susceptible-Exposed-Infected-Recovered (SEIR)
epidemiological model with stochastic forcing.  As previously mentioned, we
could easily consider a SEIR-type model where the exposed class was modeled
using hundreds of compartments.  Since the analysis is similar, we consider
the simpler standard SEIR model to demonstrate the power of our method.  Section~\ref{sec:SEIR-model}
provides a complete description of this model.  Section~\ref{sec:Dcma} describes how to
transform the deterministic SEIR system to a new system that satisfies the
spectral requirements needed to apply the center manifold theory.  After the
theory is used to find the evolution equations that describe the dynamics on
the center manifold, we show in Sec.~\ref{sec:Incorrect_proj} how the reduced
model that is found using the
deterministic result incorrectly projects the noise onto the center manifold.
Section~\ref{sec:correct_proj} demonstrates the use of a stochastic normal
form coordinate transform to correctly project the noise onto the stochastic
center manifold.  A discussion section and the conclusions are contained
respectively in Sec.~\ref{sec:disc} and Sec.~\ref{sec:conc}.

\section{The SEIR model for epidemics}
\label{sec:SEIR-model}

We begin by describing the stochastic version of the SEIR model found
in Ref.~\onlinecite{ss83}. We assume that a given population may be divided into the following four classes which evolve in time:
\begin{enumerate}
\item Susceptible class, $s(t)$, consists of those individuals who may contract the  disease.
\item Exposed class, $e(t)$, consists of those individuals who have been infected by  the disease but are not yet infectious.
\item Infectious class, $i(t)$, consists of those individuals who are capable of  transmitting the disease to susceptible individuals.
\item Recovered class, $r(t)$, consists of those individuals who are immune to the disease.
\end{enumerate}
Furthermore, we assume that the total population size, denoted as $N$, is
constant and can be normalized to $S(t)+E(t)+I(t)+R(t)=1$, where $S(t)=s(t)/N$, $E(t)=e(t)/N$,
$I(t)=i(t)/N$, and $R(t)=r(t)/N$.  Therefore, the population class variables
$S$, $E$, $I$, and $R$ represent fractions of the total population. The governing equations for the stochastic SEIR model are 
\begin{subequations}
\begin{flalign}
&\dot{S}(t)=\mu -\beta I(t)S(t) -\mu S(t) +\sigma_1\phi_1(t),\label{e:Sdot}\\
&\dot{E}(t)=\beta I(t)S(t) -(\alpha
+\mu)E(t)+\sigma_2\phi_2(t),\label{e:Edot}\\
&\dot{I}(t)=\alpha E(t) -(\gamma +\mu)I(t)+\sigma_3\phi_3(t),\label{e:Idot}\\
&\dot{R}(t)=\gamma I(t) -\mu R(t)+\sigma_4\phi_4(t),\label{e:Rdot}
\end{flalign}
\end{subequations}
where $\sigma_i$ is the standard deviation of the noise intensity
$D_i=\sigma_i^2/2$. Each of the noise terms, $\phi_i$, describes a stochastic,
Gaussian white force that is characterized by the correlation functions
\begin{subequations}
\begin{flalign}
&\langle\phi_i(t)\rangle=0,\label{e:mean}\\
&\langle\phi_i(t)\phi_j(t^{\prime})\rangle=\delta(t-t^{\prime})\delta_{ij}.\label{e:corr}
\end{flalign}
\end{subequations}

Additionally, $\mu$ represents a constant birth and death rate, $\beta$ is the
contact rate, $\alpha$ is the rate of infection, so that $1/\alpha$ is the
mean latency period, and $\gamma$ is the rate of recovery, so that $1/\gamma$
is the mean infectious period. Although the contact rate $\beta$ could be
given by a time-dependent function (e.g. due to seasonal fluctuations), for
simplicity, we assume $\beta$ to be constant. Throughout this article, we use
the following parameter values: $\mu =0.02 ({\rm  year})^{-1}$, $\beta=1575.0
({\rm year})^{-1}$, $\alpha =1/0.0279 ({\rm  year})^{-1}$, and $\gamma =1/0.01
({\rm year})^{-1}$.  Disease parameters correspond to typical measles
values~\cite{ss83,bisc02}.  Note that any other biologically meaningful
parameters may be used as long as the basic reproductive rate
$R_0=\alpha\beta/[(\alpha +\mu)(\gamma +\mu)] > 1$.  The interpretation of $R_0$ is the number of secondary cases produced by a single infectious individual in
a population of susceptibles in one infectious period.

As a first approximation of stochastic effects, we have considered additive
noise.  This type of noise may result from migration into and away from the
population being considered~\cite{BvdDW08}.  Since it is difficult to estimate
fluctuating migration rates~\cite{bfg02}, it is appropriate to treat migration as an
arbitrary external noise source.  Also, fluctuations in the birth rate
manifest itself as additive noise.  Furthermore, as we are not
interested in extinction events in this article, it is not necessary to use
multiplicative noise.  In general, for the problem considered here, it is possible that a rare
event in the tail of the noise distribution may cause one or more of the $S$,
$E$, and $I$ components of the
solution to become negative.  In this article, we will always assume that the
noise is sufficiently small 
so that a solution remains positive for a long
enough time to gather sufficient statistics.  Even though it is difficult to
accurately estimate the appropriate noise level from real data, our choices of
noise intensity lie within the huge confidence intervals computed in Ref.~\onlinecite{bfg02}.
The case for multiplicative
noise will be considered in a separate paper.

Although $S+E+I+R=1$ in the deterministic system, one should note that the dynamics of
the stochastic SEIR system will not necessarily have all of the components sum
to unity.  However, since the noise has zero mean, the total population will
remain close to unity on average.  Therefore, we assume that the dynamics  are
sufficiently described by Eqs.~(\ref{e:Sdot})-(\ref{e:Idot}).  It should be
noted that even if $E(t)+I(t)=0$ for some $t$, the noise allows for the
reemergence of the epidemic.    

\section{Deterministic center manifold analysis}\label{sec:Dcma}
One way to reduce the dimension of a system of equations is through the use of
deterministic center manifold theory.  In
general, a nonlinear vector field can be transformed so that the linear part (Jacobian)
of the vector field has block diagonal form where the first matrix block has
eigenvalues with positive real part, the second matrix block has eigenvalues
with negative real part, and the third matrix block has eigenvalues with zero
real part~\cite{hirsma74,wig90}.  These blocks are associated respectively with the unstable
eigenspace, the stable eigenspace, and the center eigenspace.  If we suppose
that there are no eigenvalues with positive real part, then orbits will
rapidly decay to the center eigenspace.

In order to make use of the center manifold theory, we must transform Eqs.~(\ref{e:Sdot})-(\ref{e:Idot}) to a new system of equations that
has the necessary spectral structure.  The theory will allow us to find an invariant center manifold passing through
the fixed point to which we can restrict the transformed system.  Details
regarding the transformation can be found in Sec.~\ref{sec:Trans}, and the computation
of the center manifold can be found in Sec.~\ref{sec:CMComp}.

\subsection{Transformation of the SEIR model}\label{sec:Trans}
Our analysis begins by considering the governing equations for the stochastic SEIR model given
by Eqs.~(\ref{e:Sdot})-(\ref{e:Idot}).  We neglect the
$\sigma_i\phi_i(t)$ terms in Eqs.~(\ref{e:Sdot})-(\ref{e:Idot}) so that we are
considering the deterministic SEIR system.  This deterministic system has two
fixed points. The first fixed point is 
\begin{equation}
\label{e:fixed_pt_mort_df}
(S_e,E_e,I_e)=(1,0,0),
\end{equation}
and corresponds to a disease free, or extinct, equilibrium state. The second
fixed point corresponds to an endemic state and is given by
\begin{equation}
\label{e:fixed_pt_mort}
(S_0,E_0,I_0)=\left (
{\frac {  \left( \gamma+\mu   \right)  \left( \alpha+\mu  \right) }{\beta \alpha}},
{\frac {\mu   }{\alpha+\mu }}-{\frac {\mu   \left( \gamma+\mu  \right) }{\alpha \beta}} ,
{\frac {\mu \alpha}{ \left( \gamma+\mu  \right)  \left( \alpha+\mu  \right) }}-{\frac {\mu  }{\beta}}
  \right ). 
\end{equation}

To ease the analysis, we define a new set of variables, $\bar{S}$, $\bar{E}$, and $\bar{I}$, as
$\bar{S}(t)=S(t)-S_0$, $\bar{E}(t)=E(t)-E_0$, and $\bar{I}(t)=I(t)-I_0$.
These new variables are substituted into Eqs.~(\ref{e:Sdot})-(\ref{e:Idot}). 

Then, treating $\mu$ as a small parameter, we rescale time by letting $t=\mu\tau$.
We may then introduce the following rescaled parameters:  $\alpha=\alpha_0/\mu$ and $\gamma=\gamma_0/\mu$,
where $\alpha_0$ and $\gamma_0$ are $\mathcal{O}(1)$.  The inclusion of the
parameter $\mu$ as a new state variable means that the terms in our rescaled
system which contain $\mu$ are now nonlinear terms.  Furthermore, the system
is augmented with the auxiliary equation $\frac{d\mu}{d\tau}=0$.  The addition
of this auxiliary equation contributes an extra simple zero eigenvalue to the
system and adds one new center direction that has trivial dynamics.  The shifted
and rescaled, augmented system of equations is 
\begin{subequations}
\begin{flalign}
\frac{d\bar{S}}{d\tau}=& -\beta\mu \bar{I}\bar{S} - \frac{(\alpha_0 +\mu^2)(\gamma_0
 +\mu^2)}{\alpha_0}\bar{I} - \frac{\alpha_0\mu^3 \beta}{(\alpha_0 +\mu^2)(\gamma_0 +\mu^2)}\bar{S},\label{e:Sbardot_mort}\\
\frac{d\bar{E}}{d\tau}=&\beta\mu \bar{I}\bar{S} + \frac{(\alpha_0 +\mu^2)(\gamma_0
 +\mu^2)}{\alpha_0}\bar{I} +\frac{\mu^2[\alpha_0\beta\mu -(\alpha_0 +\mu^2)(\gamma_0
 +\mu^2)]}{(\alpha_0 +\mu^2)(\gamma_0 +\mu^2)}\bar{S} - (\alpha_0 +\mu^2) \bar{E},\label{e:Ebardot_mort}\\
\frac{d\bar{I}}{d\tau}=&\alpha_0 \bar{E} - (\gamma_0 +\mu^2) \bar{I},\label{e:Ibardot_mort}\\
\frac{d\mu}{d\tau}=&0\label{e:mudot},
\end{flalign}
\end{subequations}
where the endemic fixed point is now located at the origin.

The Jacobian of Eqs.~(\ref{e:Sbardot_mort})-(\ref{e:mudot}) is computed to
zeroth-order in $\mu$ and is evaluated
at the origin.  Ignoring the $\mu$ components, the Jacobian has only two linearly independent eigenvectors. Therefore, the
Jacobian is not diagonalizable.  However, it is possible to
transform Eqs.~(\ref{e:Sbardot_mort})-(\ref{e:Ibardot_mort}) to a block diagonal
form with the eigenvalue structure that is needed to use center manifold
theory.  We use a transformation matrix, ${\bf P}$, consisting of the
two linearly independent eigenvectors of the Jacobian along with a third
vector chosen to be linearly independent.  There are many choices for this
third vector; our choice is predicated on keeping the vector as simple as
possible.  This transformation matrix is given as 
\begin{equation}
{\bf P} = \left [ \begin{array}{cccccccc}
1 & & & 1 & & & &0 \\
& & & & & & &\\
-\frac{\alpha_0+\gamma_0}{\gamma_0} & & & 0 & & & &0 \\
& & & & & & &\\
\frac{\alpha_0+\gamma_0}{\gamma_0} & & & 0 & & & &1  \\
 \end{array} \right ].
\end{equation}
Using the fact that
$(\bar{S},\bar{E},\bar{I})^T = {\bf P}\cdot (U,V,W)^T$, then the transformation matrix leads to the following definition of new variables, $U$, $V$, and $W$:
\begin{subequations}
\begin{flalign}
&U = \frac{-\gamma_0}{\alpha_0 + \gamma_0}\bar{E},\label{e:U}\\
&V = \bar{S} + \frac{\gamma_0}{\alpha_0 + \gamma_0}\bar{E},\label{e:V}\\
&W = \bar{I} + \bar{E}.\label{e:W}
\end{flalign}
\end{subequations}

The application of the transformation matrix to
Eqs.~(\ref{e:Sbardot_mort})-(\ref{e:Ibardot_mort}) leads to the transformed
evolution equations given by 
\begin{subequations}
\begin{flalign}
\frac{dU}{d\tau}=& -\alpha_0 U+{\frac {\mu^2 \left( \gamma_0 V-\alpha_0 U \right) }{\alpha_0+\gamma_0}}-{\frac { \left( \gamma_0+\mu^2 \right) \left( \alpha_0+\mu^2 \right) \left[ \left( \alpha_0+\gamma_0 \right) U +\gamma_0 W \right] }{\alpha_0 \left( \alpha_0+\gamma_0 \right) }}- &&\nonumber \\
& \frac{\mu \beta}{\alpha_0+\gamma_0}  \left( \gamma_0 W+\left( \alpha_0+\gamma_0 \right) U +{\frac {\mu^2\alpha_0 \gamma_0}{ \left( \gamma_0+\mu^2 \right) \left( \alpha_0+\mu^2 \right) }} \right) \left( U+V \right) 
,\label{e:dU_mort}
\end{flalign}
\begin{flalign}
\frac{dV}{d\tau}=& \alpha_0 U-{\frac {\mu^2 \left( \gamma_0 V -\alpha_0 U \right) }{\alpha_0+\gamma_0}}-{\frac { \left( \gamma_0+\mu^2 \right) \left( \alpha_0+\mu^2 \right) \left[ \left( \alpha_0+\gamma_0 \right) U +\gamma_0 W \right] }{\gamma_0 \left( \alpha_0+\gamma_0 \right) }}- &&\nonumber \\
& \frac{\mu \beta \alpha_0}{ \gamma_0 \left( \alpha_0+\gamma_0 \right)} \left( \gamma_0 W+\left( \alpha_0+\gamma_0 \right) U +{\frac {\mu^2\alpha_0 \gamma_0}{ \left( \gamma_0+\mu^2 \right) \left( \alpha_0+\mu^2 \right) }} \right) \left( U+V \right)
,\label{e:dV_mort}
\end{flalign}
\begin{flalign}
\frac{dW}{d\tau}=& -\alpha_0 U - \left( \gamma_0+\mu^2 \right) \left( U+W \right) +{\frac { \left( \gamma_0+\mu^2 \right) \left( \alpha_0+\mu^2 \right) \left[ \left( \alpha_0+\gamma_0 \right) U +\gamma_0 W \right] }{\alpha_0 \gamma_0}}- &&\nonumber \\
& \mu^2V + \frac{\mu \beta}{\gamma_0} \left( \gamma_0 W+\left( \alpha_0+\gamma_0 \right) U +{\frac {\mu^2\alpha_0 \gamma_0}{ \left( \gamma_0+\mu^2 \right) \left( \alpha_0+\mu^2 \right) }} \right) \left( U+V \right)
 ,\label{e:dW_mort}
\end{flalign}
\begin{flalign}
\frac{d\mu}{d\tau}=&0.&&\label{e:dmu_mort} 
\end{flalign}
\end{subequations}

\subsection{Center manifold equation}\label{sec:CMComp}
The Jacobian of Eqs.~(\ref{e:dU_mort})-(\ref{e:dmu_mort}) to zeroth-order in
$\mu$ and evaluated at the origin is
\begin{equation}
\left [ \begin{array}{cc|ccccccccc}
-(\alpha_0+\gamma_0) & & & &0 & & & -\frac{\gamma_0^2}{(\alpha_0+\gamma_0)}&
& & 0 \\
& & & & & & & & & &\\
\hline
& & & & & & & & & &\\
0 & & & &0 & & & -\frac{\alpha_0\gamma_0}{(\alpha_0+\gamma_0)}& & & 0 \\
& & & & & & & & & &\\
0 & & & &0 & & & 0 & & & 0 \\
& & & & & & & & & &\\
0 & & & &0 & & & 0 & & & 0 \end{array} \right ],
\end{equation} 
which shows that Eqs.~(\ref{e:dU_mort})-(\ref{e:dmu_mort}) may be rewritten in the form
\begin{flalign}
&\frac{d{\bf x}}{d\tau}= {\bf A}{\bf x} + {\bf f}({\bf x},{\bf  y},\mu),\label{e:CMformx}\\
&\frac{d{\bf y}}{d\tau}= {\bf B}{\bf y} + {\bf g}({\bf x},{\bf  y},\mu),\label{e:CMformy}\\
&\frac{d\mu}{d\tau}=0,
\end{flalign}
where ${\bf x}=(U)$, ${\bf y}=(V,W)$, ${\bf A}$ is a constant matrix with
eigenvalues that have negative real parts, ${\bf B}$ is a constant matrix with
eigenvalues that have zero real parts, and ${\bf f}$ and ${\bf g}$ are
nonlinear functions in ${\bf x}$, ${\bf y}$ and $\mu$.  In particular,
\begin{equation}
{\bf A}=\left [ \begin{array}{c}
-(\alpha_0 +\gamma_0)\end{array} \right ], \,\,\,\,\,\,\,\,\,\,\,\,\,\,\, {\bf B}=\left [
\begin{array}{cccc}
0 &&& -\frac{\alpha_0\gamma_0}{(\alpha_0 +\gamma_0)}\\
&&&\\
0 &&& 0\end{array} \right ].
\end{equation}

 Therefore, the system will rapidly collapse onto a lower-dimensional manifold
 given by center manifold theory~\cite{car81}. Furthermore, we know that the center manifold is given by
\begin{equation}
\label{e:CM}
U=h(V,W,\mu),
\end{equation}
where $h$ is an unknown function.

Substitution of Eq.~(\ref{e:CM}) into Eq.~(\ref{e:dU_mort}) leads to the
following center manifold condition:
\begin{flalign}
\label{e:CMcond}
&\frac{\partial h(V,W,\mu)}{\partial V}\frac{dV}{d\tau}+\frac{\partial
  h(V,W,\mu)}{\partial W}\frac{dW}{d\tau}= -\alpha_0 h(V,W,\mu)+ {\frac {\mu^2
    \left[ \gamma_0 V-\alpha_0 h(V,W,\mu) \right]
  }{\alpha_0+\gamma_0}}-&&\nonumber\\
&{\frac { \left( \gamma_0+\mu^2 \right) \left( \alpha_0+\mu^2 \right) \left[ \left( \alpha_0+\gamma_0 \right) h(V,W,\mu) +\gamma_0 W \right] }{\alpha_0 \left( \alpha_0+\gamma_0 \right) }}- \nonumber \\
& \frac{\mu \beta}{\alpha_0+\gamma_0}  \left( \gamma_0 W+\left( \alpha_0+\gamma_0 \right) h(V,W,\mu) +{\frac {\mu^2\alpha_0 \gamma_0}{ \left( \gamma_0+\mu^2 \right) \left( \alpha_0+\mu^2 \right) }} \right) \left( h(V,W,\mu)+V \right).
\end{flalign}
In general, it is not possible to solve the center manifold condition for the unknown function, $h(V,W,\mu)$. Therefore, we perform the following Taylor series expansion of $h(V,W,\mu)$ in $V$, $W$, and $\mu$:
\begin{flalign}
\label{e:TS}
h(V,W,\mu)=&h_0 + h_2 V +h_3 W +h_\mu\mu + h_{22}V^2 + h_{23}VW +h_{33}W^2
+\nonumber\\
&h_{\mu 2}\mu V + h_{\mu 3}\mu W + h_{\mu\mu}\mu^2 + \ldots,
\end{flalign} 
where $h_0$, $h_2$, $h_3$, $h_\mu$, $\ldots$ are unknown coefficients that are found by substituting the Taylor series expansion into the center manifold condition and equating terms of the same order. By carrying out this procedure using a second-order Taylor series expansion of $h$, the center manifold equation is
\begin{equation}
\label{e:CMeq}
U=-\frac{\gamma_0^2}{(\alpha_0+\gamma_0)^2}W + \mathcal{O}(\epsilon^3),
\end{equation}
where $\epsilon =|(V,W,\mu)|$ so that $\epsilon$ provides a count of the
number of $V$, $W$, and $\mu$ factors in any one term.  Substitution of Eq.~(\ref{e:CMeq}) into Eqs.~(\ref{e:dV_mort}) and~(\ref{e:dW_mort}) leads to the following reduced system of evolution equations which describe the dynamics on the center manifold: 
\begin{subequations}
\begin{flalign}
\frac{dV}{d\tau}=& -{\frac {\mu^2{\gamma_0}^2\alpha_0  W}{ \left(
      \alpha_0+\gamma_0 \right) ^{3}}} -{\frac {  \mu^{4}\alpha_0 W}{ \left(
      \alpha_0+\gamma_0 \right)^2}}  -{\frac {\gamma_0 \mu^2
    V}{\alpha_0+\gamma_0}}-\nonumber\\
&{\frac { \left( \gamma_0+\mu^2 \right)  \alpha_0 W}{\alpha_0+\gamma_0}}-
\frac{\beta {\alpha_0}^2\mu}{ \left( \alpha_0+\gamma_0
  \right)^2} \left(  W+{\frac {\mu^2 \left( \alpha_0+\gamma_0 \right) }{ \left( \gamma_0+\mu^2 \right)  \left( \alpha_0+\mu^2 \right) }} \right)  \left(  V-{\frac {{\gamma_0}^2 W}{ \left( \alpha_0+\gamma_0 \right)^2}} \right) 
,\label{e:dV_mort_red}\\
\frac{dW}{d\tau}=& {\frac {\mu^2 {\gamma_0}^2W }{ \left( \alpha_0+\gamma_0
    \right)^2}}+{\frac {\mu^{4} W}{\alpha_0+\gamma_0}}-\mu^2 V+\nonumber\\
& \frac{\beta
  \mu \alpha_0}{ \alpha_0+\gamma_0 }  \left(  W+{\frac {\mu^2 \left( \alpha_0+\gamma_0 \right) }{ \left( \gamma_0+\mu^2 \right)  \left( \alpha_0+\mu^2 \right) }} \right)  \left(  V-{\frac {{\gamma_0}^2 W}{ \left( \alpha_0+\gamma_0 \right)^2}} \right) 
.\label{e:dW_mort_red} 
\end{flalign}
\end{subequations}

\section{Incorrect projection of the noise onto the stochastic center
 manifold}\label{sec:Incorrect_proj}
\subsection{Transformation of the stochastic SEIR model}
We now return to the stochastic SEIR system given by Eqs.~(\ref{e:Sdot})-(\ref{e:Idot}). The shift of the fixed point to the origin will not have any effect on the noise terms, so that the stochastic version of the shifted equations is
\begin{subequations}
\begin{flalign}
&\dot{\bar{S}}(t)= -\beta \bar{I}\bar{S} - \frac{(\alpha +\mu)(\gamma
 +\mu)}{\alpha}\bar{I} - \frac{\alpha\mu \beta}{(\alpha +\mu)(\gamma +\mu)}\bar{S}+\sigma_1\phi_1(t),\label{e:Sbardot_mort_stoch}\\
&\dot{\bar{E}}(t)=\beta \bar{I}\bar{S} + \frac{(\alpha +\mu)(\gamma
 +\mu)}{\alpha}\bar{I} +\frac{\mu[\alpha\beta-(\alpha +\mu)(\gamma
 +\mu)]}{(\alpha +\mu)(\gamma +\mu)}\bar{S} - (\alpha +\mu) \bar{E}+\sigma_2\phi_2(t),\label{e:Ebardot_mort_stoch}\\
&\dot{\bar{I}}(t)=\alpha \bar{E} - (\gamma +\mu) \bar{I}+\sigma_3\phi_3(t).\label{e:Ibardot_mort_stoch}
\end{flalign}
\end{subequations}

As Eqs.~(\ref{e:Sbardot_mort_stoch})-(\ref{e:Ibardot_mort_stoch}) are
transformed using Eqs.~(\ref{e:U})-(\ref{e:W}), the $\alpha=\alpha_0/\mu$ scaling, the
$\gamma=\gamma_0/\mu$ scaling, and the $t=\mu\tau$ time scaling, the noise
also is
scaled so that the stochastic, transformed equations are given by 
\begin{subequations}
\begin{flalign}
\frac{dU}{d\tau}= &-\alpha_0 U+{\frac {\mu^2 \left( \gamma_0 V-\alpha_0 U \right) }{\alpha_0+\gamma_0}}-{\frac { \left( \gamma_0+\mu^2 \right) \left( \alpha_0+\mu^2 \right) \left[ \left( \alpha_0+\gamma_0 \right) U +\gamma_0 W \right] }{\alpha_0 \left( \alpha_0+\gamma_0 \right) }}- \nonumber \\
& \frac{\mu \beta}{\alpha_0+\gamma_0}  \left( \gamma_0 W+\left( \alpha_0+\gamma_0 \right) U +{\frac {\mu^2\alpha_0 \gamma_0}{ \left( \gamma_0+\mu^2 \right) \left( \alpha_0+\mu^2 \right) }} \right) \left( U+V \right) 
+\sigma_4\phi_4
,\label{e:dU_mort_stoch}\\
\frac{dV}{d\tau}=& \alpha_0 U-{\frac {\mu^2 \left( \gamma_0 V -\alpha_0 U \right) }{\alpha_0+\gamma_0}}-{\frac { \left( \gamma_0+\mu^2 \right) \left( \alpha_0+\mu^2 \right) \left[ \left( \alpha_0+\gamma_0 \right) U +\gamma_0 W \right] }{\gamma_0 \left( \alpha_0+\gamma_0 \right) }}- \nonumber \\
& \frac{\mu \beta \alpha_0}{ \gamma_0 \left( \alpha_0+\gamma_0 \right)} \left( \gamma_0 W+\left( \alpha_0+\gamma_0 \right) U +{\frac {\mu^2\alpha_0 \gamma_0}{ \left( \gamma_0+\mu^2 \right) \left( \alpha_0+\mu^2 \right) }} \right) \left( U+V \right)
+\sigma_5\phi_5,\label{e:dV_mort_stoch}\\
\frac{dW}{d\tau}=&   -\alpha_0 U - \left( \gamma_0+\mu^2 \right) \left( U+W \right) +{\frac { \left( \gamma_0+\mu^2 \right) \left( \alpha_0+\mu^2 \right) \left[ \left( \alpha_0+\gamma_0 \right) U +\gamma_0 W \right] }{\alpha_0 \gamma_0}}- \nonumber \\
& \mu^2V + \frac{\mu \beta}{\gamma_0} \left( \gamma_0 W+\left( \alpha_0+\gamma_0 \right) U +{\frac {\mu^2\alpha_0 \gamma_0}{ \left( \gamma_0+\mu^2 \right) \left( \alpha_0+\mu^2 \right) }} \right) \left( U+V \right)
+\sigma_6\phi_6,\label{e:dW_mort_stoch}
\end{flalign}
\end{subequations}
where 
\begin{subequations}
\begin{flalign}
\sigma_4\phi_4 = &-\frac{\mu\gamma_0 }{\alpha_0
 +\gamma_0}\sigma_2\phi_2,\label{e:sig4phi4}\\
\sigma_5\phi_5 = & \mu\sigma_1\phi_1 + \frac{\mu\gamma_0}{\alpha_0
  +\gamma_0}\sigma_2\phi_2,\label{e:sig5phi5}\\
\sigma_6\phi_6 = & \mu\sigma_2\phi_2 +\mu\sigma_3\phi_3.\label{e:sig6phi6}
\end{flalign}
\end{subequations}
The stochastic terms $\phi_4$, $\phi_5$, and $\phi_6$
in Eqs.~(\ref{e:dU_mort_stoch})-(\ref{e:dW_mort_stoch}) are still additive,
Gaussian noise processes.  However,
Eqs.~(\ref{e:sig4phi4})-(\ref{e:sig6phi6}) show how the transformation has
acted on the original stochastic terms $\phi_1$, $\phi_2$, and $\phi_3$ to
create new noise processes which have a variance different from that of the
original noise processes.  Also note that we have suppressed the
argument of $\phi_4$, $\phi_5$, and $\phi_6$ in
Eqs.~(\ref{e:dU_mort_stoch})-(\ref{e:dW_mort_stoch}).  The time
scaling means that these noise terms should be evaluated at $\mu\tau$.

\begin{figure}[h!]
\includegraphics[width=8.5cm,height=6.375cm]{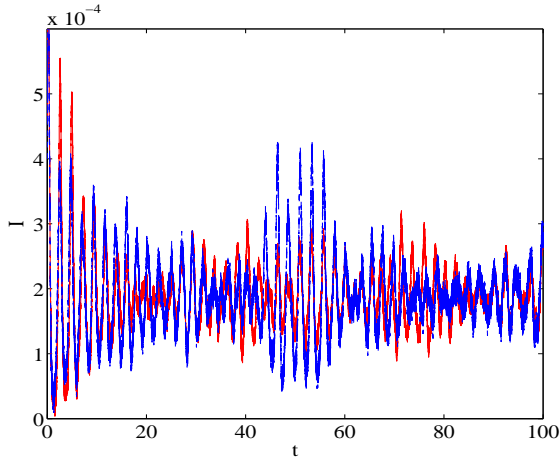}
\caption{\label{fig:TS_SEI_mort_Trans_mort_full}(Color online) Time series of the fraction of
  the population that is  infected with a disease, $I$, computed using the
  original, stochastic system of equations of the SEIR model [Eqs.~(\ref{e:Sdot})-(\ref{e:Idot})] (red, solid line), and
  computed using the transformed, stochastic system of equations of the SEIR
  model [Eqs.~(\ref{e:dU_mort_stoch})-(\ref{e:dW_mort_stoch})] (blue, dashed
  line). The standard deviation of the noise intensity used in the simulation
  is $\sigma_i =0.0005$, $i=1,\ldots,6$.}
\end{figure}

The system of equations given by Eqs.~(\ref{e:dU_mort_stoch})-(\ref{e:sig6phi6})
are an exact transformation of the system of equations given by
Eqs.~(\ref{e:Sdot})-(\ref{e:Idot}).  We numerically integrate the original, stochastic system of the SEIR model [Eqs.~(\ref{e:Sdot})-(\ref{e:Idot})] along with the transformed, stochastic
system [Eqs.~(\ref{e:dU_mort_stoch})-(\ref{e:dW_mort_stoch})] using a stochastic
fourth-order Runge-Kutta scheme with a constant time step size. The original
system is solved for $S$, $E$, and $I$, while the transformed system is solved
for $U$, $V$, and $W$. In the latter case, once the values of $U$, $V$, and $W$
are known, we compute the values of $\bar{S}$, $\bar{E}$, and $\bar{I}$ using
the transformation given by Eqs.~(\ref{e:U})-(\ref{e:W}). We shift $\bar{S}$,
$\bar{E}$, and $\bar{I}$ respectively by $S_0$, $E_0$, and $I_0$ to find the
values of $S$, $E$, and $I$.

Figure~\ref{fig:TS_SEI_mort_Trans_mort_full} compares the time series of the
fraction of the population that is infected with a disease, $I$, computed
using the original, stochastic system of equations of the SEIR model 
with the time series of $I$ computed using the transformed,
stochastic system of equations of the SEIR model. 
 
Although the two time series shown in Fig.~\ref{fig:TS_SEI_mort_Trans_mort_full} generally agree very well, there is some discrepancy.  This
discrepancy is due to the fact that the noise processes $\sigma_4\phi_4$,
$\sigma_5\phi_5$, and $\sigma_6\phi_6$ of the transformed system are new,
independent noise processes with different variance than the $\sigma_1\phi_1$,
$\sigma_2\phi_2$, and $\sigma_3\phi_3$ noise processes found in the original
system.  If we carefully take the noise realization from the original system,
transform this noise using Eqs.~(\ref{e:sig4phi4})-(\ref{e:sig6phi6}), and use
this realization to solve the transformed system, then the two solutions would
be identical.

\subsection{Reduction of the stochastic SEIR model}\label{sec:ipcm_mort}

It is tempting to consider the reduced stochastic model found by substitution
of Eq.~(\ref{e:CMeq}) into Eqs.~(\ref{e:dV_mort_stoch}) and~(\ref{e:dW_mort_stoch}), so that one has the following stochastic evolution equations (that hopefully describe the dynamics on the stochastic center manifold):
\begin{subequations}
\begin{flalign}
\frac{dV}{d\tau}=& -{\frac {\mu^2{\gamma_0}^2\alpha_0  W}{ \left( \alpha_0+\gamma_0 \right) ^{3}}} -{\frac {  \mu^{4}\alpha_0 W}{ \left( \alpha_0+\gamma_0 \right)^2}}  -{\frac {\gamma_0 \mu^2 V}{\alpha_0+\gamma_0}}-{\frac { \left( \gamma_0+\mu^2 \right)  W \alpha_0}{\alpha_0+\gamma_0}}- \nonumber \\
& \frac{\beta {\alpha_0}^2\mu}{ \left( \alpha_0+\gamma_0 \right)^2}  \left(  W+{\frac {\mu^2 \left( \alpha_0+\gamma_0 \right) }{ \left( \gamma_0+\mu^2 \right)  \left( \alpha_0+\mu^2 \right) }} \right)  \left(  V-{\frac {{\gamma_0}^2 W}{ \left( \alpha_0+\gamma_0 \right)^2}} \right) 
+\sigma_5\phi_5,\label{e:dV_mort_red_stoch}\\
\frac{dW}{d\tau}=& {\frac {\mu^2 {\gamma_0}^2W }{ \left( \alpha_0+\gamma_0 \right)^2}}+{\frac {\mu^{4} W}{\alpha_0+\gamma_0}}-\mu^2 V+  \nonumber \\
&  \frac{\beta \mu \alpha_0}{ \alpha_0+\gamma_0 }  \left(  W+{\frac {\mu^2 \left( \alpha_0+\gamma_0 \right) }{ \left( \gamma_0+\mu^2 \right)  \left( \alpha_0+\mu^2 \right) }} \right)  \left(  V-{\frac {{\gamma_0}^2 W}{ \left( \alpha_0+\gamma_0 \right)^2}} \right) + \sigma_6\phi_6.\label{e:dW_mort_red_stoch} 
\end{flalign}
\end{subequations}

One should note that Eqs.~(\ref{e:dV_mort_red_stoch})
and~(\ref{e:dW_mort_red_stoch}) also can be found by na\"{i}vely adding the
stochastic terms to the reduced system of evolution equations for the
deterministic problem [Eqs.~(\ref{e:dV_mort_red}) and~(\ref{e:dW_mort_red})].  This type of stochastic center manifold reduction has been done for the case
of discrete noise~\cite{bisc02}.  Additionally, in many other fields
(e.g. oceanography, solid mechanics, fluid mechanics), researchers have
performed stochastic model reduction using a Karhunen-Lo\`{e}ve expansion
(principal component analysis, proper orthogonal decomposition)~\cite{doghre07,vewaka08}.
However, this linear projection does not properly capture the nonlinear
effects.  Furthermore, one must subjectively choose the number of modes needed
for the expansion.  Therefore, even though the solution to the reduced model
found using this technique may have the correct statistics, individual
solution realizations will not agree with the original, complete solution.  

We will show that Eqs.~(\ref{e:dV_mort_red_stoch})
and~(\ref{e:dW_mort_red_stoch}) do not contain the correct projection of the
noise onto the center manifold.  Therefore, when solving the reduced system,  one does not obtain the correct solution. Such errors in stochastic  epidemic modeling impact the prediction of disease outbreak when modeling  the spread of a disease in a population. 

\begin{figure}[h!]
\includegraphics[width=8.5cm,height=5cm]{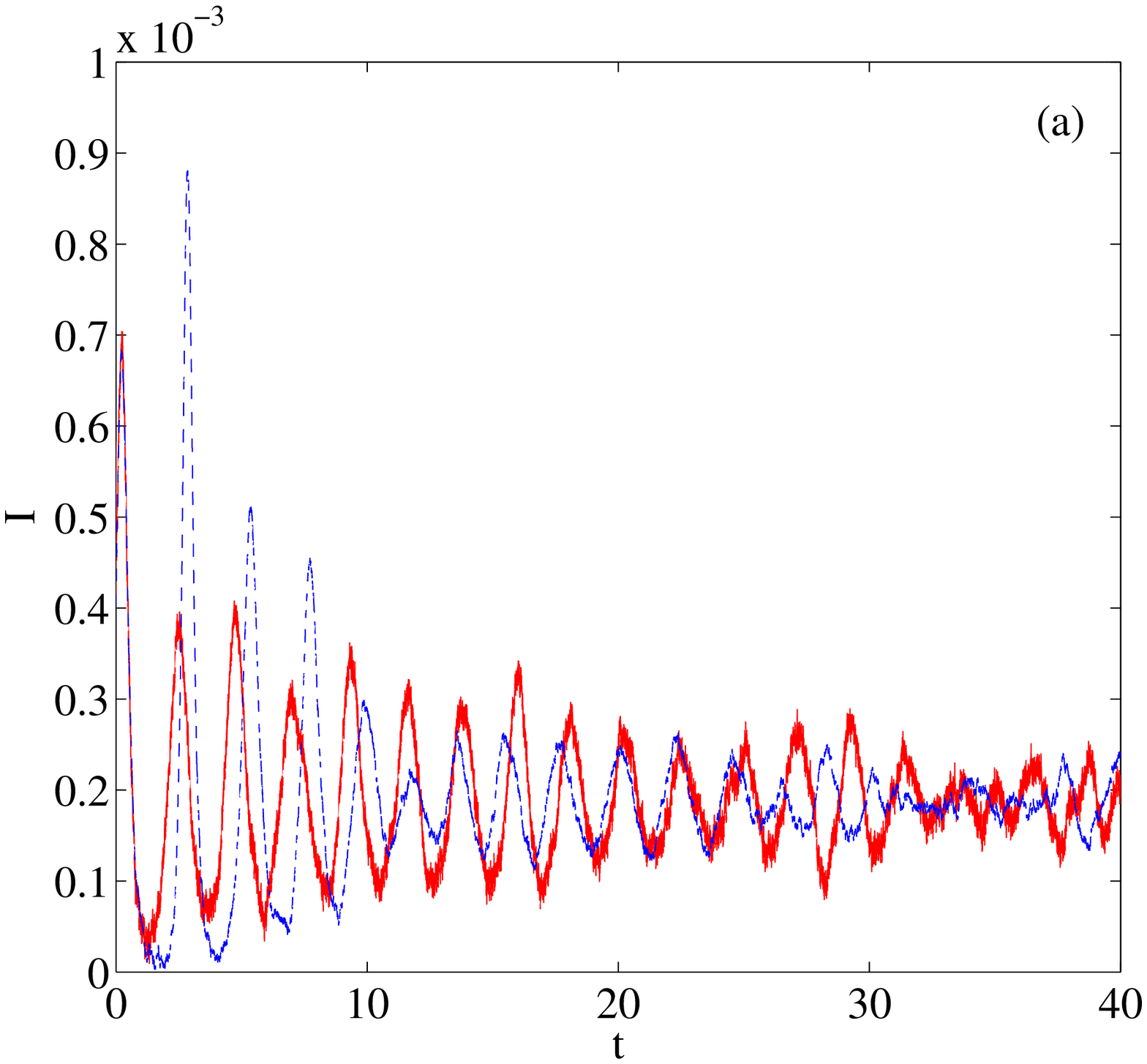}\\
\includegraphics[width=8.5cm,height=5cm]{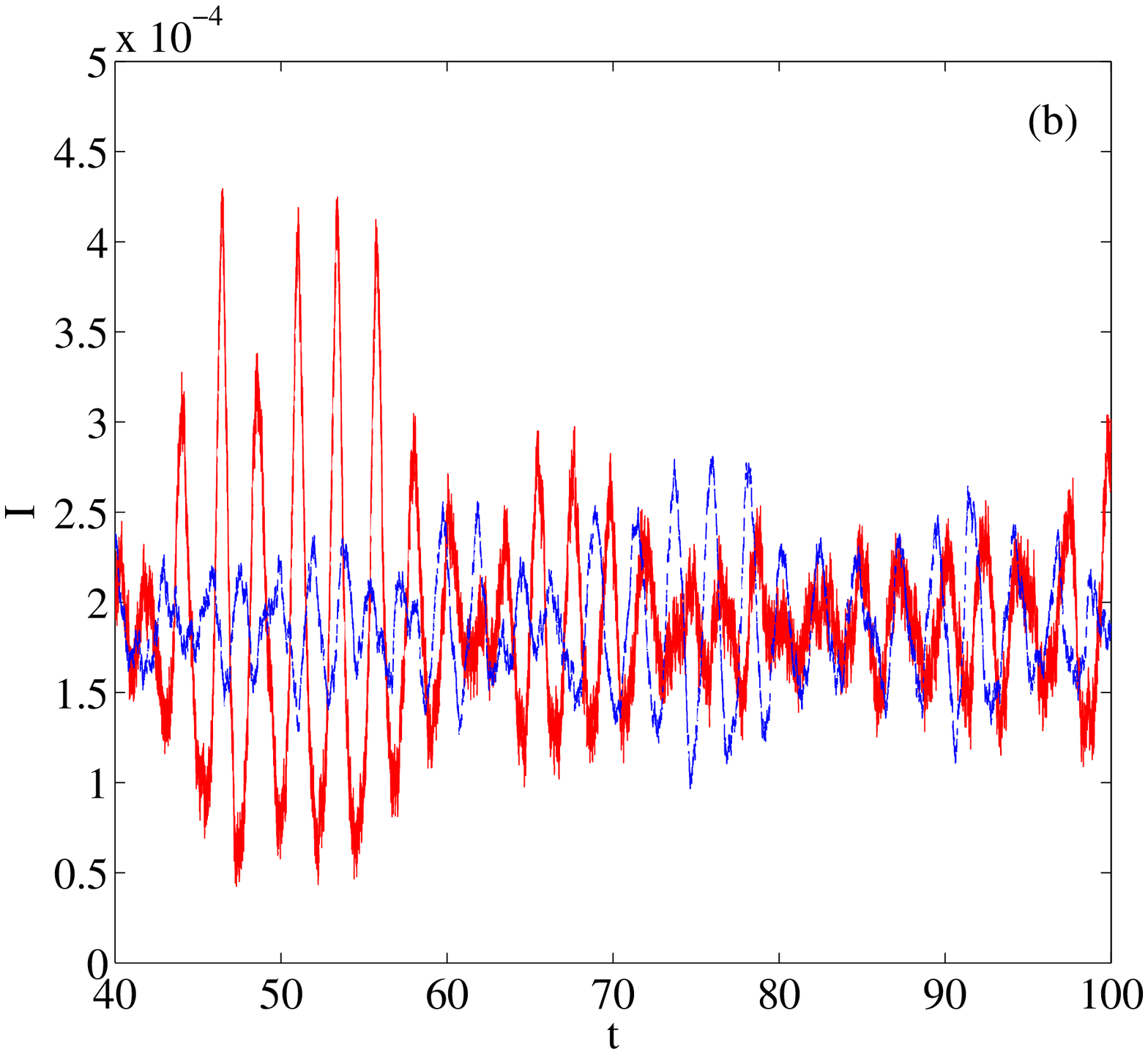}
\caption{\label{fig:TS_full_reducedCM}(Color online) Time series of the fraction of the
  population that is  infected with a disease, $I$, computed using the
  complete, stochastic system of transformed equations of the SEIR model
  [Eqs.~(\ref{e:dU_mort_stoch})-(\ref{e:dW_mort_stoch})] (red, solid line),
  and computed using the reduced system of equations of the SEIR model that is based on the
  deterministic center manifold with a replacement of the noise terms
  [Eqs.~(\ref{e:dV_mort_red_stoch}) and~(\ref{e:dW_mort_red_stoch})] (blue,
  dashed line). The standard deviation of the noise intensity used in the
  simulation is $\sigma_i =0.0005$, $i=4,5,6$.  The time series is shown for (a)
  $t=0$ to $t=40$, and for (b) $t=40$ to $t=100$.}
\end{figure}

Using the same numerical scheme previously described, we numerically integrate
the complete, stochastic system of transformed equations of the SEIR model [Eqs.~(\ref{e:dU_mort_stoch})-(\ref{e:dW_mort_stoch})] along
with the reduced system of equations that is based on the deterministic center manifold
with a replacement of the noise terms [Eqs.~(\ref{e:dV_mort_red_stoch})
and~(\ref{e:dW_mort_red_stoch})].  The complete system is solved for $U$, $V$, and $W$, while the reduced system is solved for $V$ and $W$. In the latter case, $U$ is computed using the center manifold equation given by Eq.~(\ref{e:CMeq}). Once the values of $U$, $V$, and $W$ are known, we compute the values of $\bar{S}$, $\bar{E}$, and $\bar{I}$ using the transformation given by Eqs.~(\ref{e:U})-(\ref{e:W}). We shift $\bar{S}$, $\bar{E}$, and $\bar{I}$ respectively by $S_0$, $E_0$, and $I_0$ to find the values of $S$, $E$, and $I$.

Figures~\ref{fig:TS_full_reducedCM}(a)-(b) compares the time series of the fraction of
the population that is infected with a disease, $I$, computed using the
complete, stochastic system of transformed equations of the SEIR model
[Eqs.~(\ref{e:dU_mort_stoch})-(\ref{e:dW_mort_stoch})] with the time series of
$I$ computed using the reduced system of equations of the SEIR model that is based on the
deterministic center manifold with a replacement of the noise terms
[Eqs.~(\ref{e:dV_mort_red_stoch}) and~(\ref{e:dW_mort_red_stoch})].
Figure~\ref{fig:TS_full_reducedCM}(a) shows the initial transients, while
Fig.~\ref{fig:TS_full_reducedCM}(b) shows the time series after the transients have
decayed.  One can see that the solution computed using the reduced system quickly
becomes out of phase with the solution of the complete system.  Use of this
reduced system would lead to an incorrect prediction of the initial disease
outbreak.  Additionally, the predicted amplitude of the initial outbreak would
be incorrect.  The poor agreement, both in phase and amplitude, between the
two solutions continues for long periods of time as seen in Fig.~\ref{fig:TS_full_reducedCM}(b).  We also have
computed the cross-correlation of the two time series shown in
Fig.~\ref{fig:TS_full_reducedCM}(a)-(b) to be approximately 0.34.  Since the cross-correlation measures the
similarity between the two time series, this low value quantitatively suggests
poor agreement between the two solutions.

\begin{figure}[h!]
\includegraphics[width=8.5cm,height=7cm]{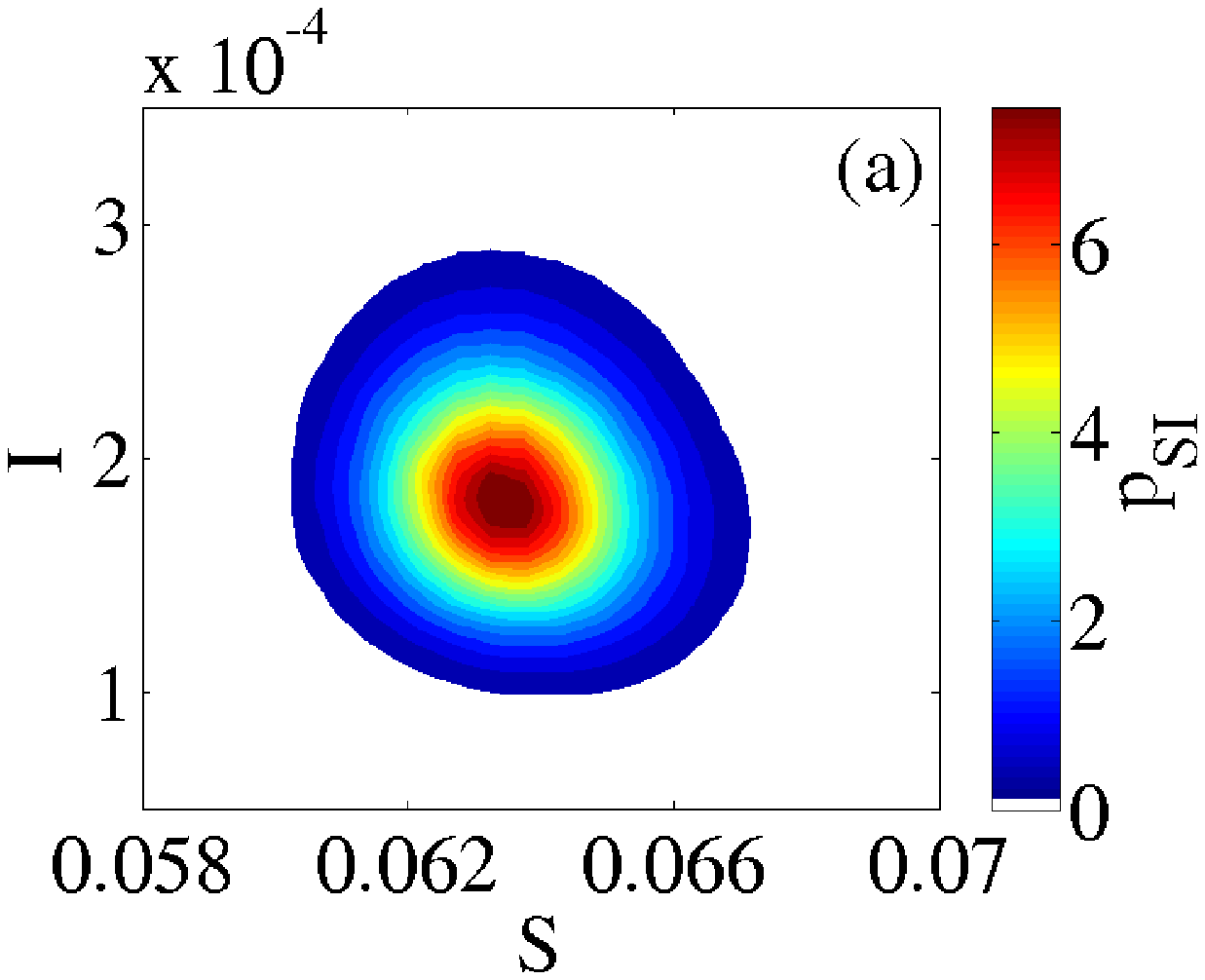}
\includegraphics[width=8.5cm,height=7cm]{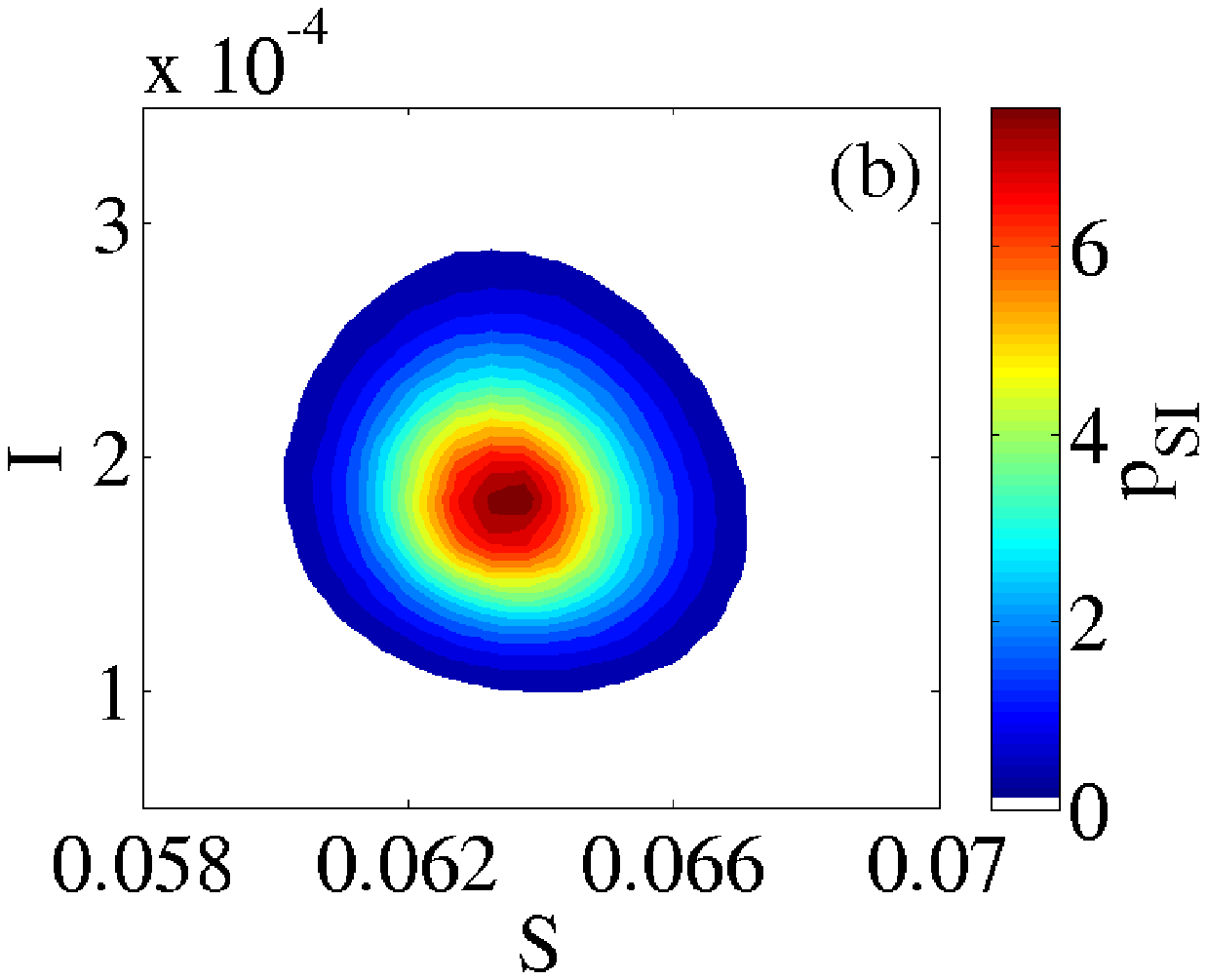}
\caption{\label{fig:hist_full_reducedCM} (Color online) Histogram of probability density,
  $p_{SI}$ of the $S$ and $I$ values found using (a) the complete, stochastic
  system of transformed equations for the SEIR model
  [Eqs.~(\ref{e:dU_mort_stoch})-(\ref{e:dW_mort_stoch})], and (b)  the reduced
  system of equations of the SEIR model that is based on the  deterministic center manifold
  with a replacement of the noise terms  [Eqs.~(\ref{e:dV_mort_red_stoch}) and~(\ref{e:dW_mort_red_stoch})]. The  histograms are created using $100$ years worth of time series (starting with year $40$) for 500 realizations, and the color-bar values have been normalized by $10^{-3}$.}
\end{figure}

Using the same systems of transformed equations, we compute $140$ years worth
of time series for $500$ realizations. Ignoring the first $40$ years of
transient solution, the data is used to create a histogram representing the
probability density, $p_{SI}$ of the $S$ and $I$
values. Figure~\ref{fig:hist_full_reducedCM}(a) shows the histogram associated
with the complete, stochastic system of transformed equations, while
Fig.~\ref{fig:hist_full_reducedCM}(b) shows the histogram associated with the
reduced system of equations with a replacement of the noise terms. The color-bar values in
Figs.~\ref{fig:hist_full_reducedCM}(a)-(b) have been normalized by $10^{-3}$.

One can see by comparing Fig.~\ref{fig:hist_full_reducedCM}(a) with
Fig.~\ref{fig:hist_full_reducedCM}(b) that the two probability distributions
qualitatively look the same. It also is possible to compare the two distributions using a quantitative measure. The Kullback-Leibler divergence, or relative entropy, measures the difference between the two probability distributions as
\begin{equation}
\label{e:KL}
d_{KL}=\sum\limits_{i,j}P_{i,j}\left|\log{\left (\frac{P_{i,j}}{Q_{i,j}}\right )}\right|,
\end{equation}
where $P_{i,j}$ refers to the $(i,j)$th component of the probability density
found using the complete, stochastic system of transformed equations
[Fig.~\ref{fig:hist_full_reducedCM}(a)], and $Q_{i,j}$ refers to the $(i,j)$th
component of the probability density found using the reduced system of
equations [Fig.~\ref{fig:hist_full_reducedCM}(b)].  In our numerical computation of the
relative entropy, we have added $10^{-10}$ to each $P_{ij}$ and $Q_{ij}$.
This eliminates the possibility of having a $Q_{ij}=0$ in the denominator of
Eq.~(\ref{e:KL}) and does not have much of an effect on the relative entropy.

If the two histograms were identical, then the relative entropy given by
Eq.~(\ref{e:KL}) would be $d_{KL}=0$. The two histograms shown in
Figs.~\ref{fig:hist_full_reducedCM}(a)-(b) have a relative entropy of
$d_{KL}=0.0391$, which means that the two histograms, while not identical, are
quantitatively very similar. However, one cannot rely entirely on the
histograms alone to say that the solutions of the complete system and the
reduced system agree. As we have seen in Figs.~\ref{fig:TS_full_reducedCM}(a)-(b), the
two solutions have differing amplitudes and are out of phase with one
another.  It is important to note that these features are not picked up by the histograms of Fig.~\ref{fig:hist_full_reducedCM}.

\section{Correct projection of the noise onto the stochastic center
 manifold}\label{sec:correct_proj}
To project the noise correctly onto the center manifold, we will derive a
normal form coordinate transform for the complete, stochastic system of
transformed equations of the SEIR model given by
Eqs.~(\ref{e:dU_mort_stoch})-(\ref{e:dW_mort_stoch}). The particular method we
use to construct the normal form coordinate transform not only reduces the
dimension of the dynamics, but also separates all of the fast processes from
all of the slow processes~\cite{rob08}.  This technique has been modified and applied to the large fluctuations of  multiscale problems~\cite{forsch09}.

Many publications~\cite{knowie83,coelti85,nam90,namlin91} discuss the simplification of a stochastic dynamical system using a stochastic normal form transformation. In some of this work~\cite{coelti85,namlin91}, anticipative noise processes appeared in the normal form transformations, but these integrals of the noise process into the future were not dealt with rigorously. 

Later, the rigorous, theoretical analysis needed to support normal form
coordinate transforms was developed in Refs.~\onlinecite{arnimk98,arn98}. The technical problem of the anticipative noise integrals also was dealt with rigorously in this work. Even later, another stochastic normal form transformation was developed~\cite{rob08}. This new method allows for the ``[removal of] anticipation ... from the slow modes with the result that no anticipation is required after the fast transients decay''(Ref.~\onlinecite{rob08}, pp. 13). The removal of anticipation leads to a simplification of the normal form. Nonetheless, this simpler normal form retains its accuracy with the original stochastic system~\cite{rob08}.
 
We shall use the method of Ref.~\onlinecite{rob08} to simplify our stochastic dynamical
system to one that emulates the long-term dynamics of the original system. The
method involves five principles, which we recapitulate here for completeness:
\begin{enumerate}
\item Avoid unbounded, secular terms in both the transformation and the  evolution equations to ensure a uniform asymptotic approximation.\label{one}
\item Decouple all of the slow processes from the fast processes to ensure a  valid long-term model.\label{two}
\item Insist that the stochastic slow manifold is precisely the transformed  fast processes coordinate being equal to zero.\label{three}
\item To simplify matters, eliminate as many as possible of the terms in the evolution equations.\label{four}
\item Try to remove all fast processes from the slow processes by avoiding as  much as possible the fast time memory integrals in the evolution equations.\label{five}
\end{enumerate}

In practice, the original stochastic system of equations (which satisfies the
necessary spectral requirements) in $\left (U,V,W\right )^T$ coordinates is
transformed to a new $\left (Y,X_1,X_2\right )^T$ coordinate system using a
near-identity stochastic coordinate transform given as
\begin{subequations}
\begin{flalign}
U=&Y+ \xi\left
 (Y,X_1,X_2,\tau\right ),\\
V=&X_1+ \eta\left (Y,X_1,X_2,\tau\right ),\\
W=&X_2+ \rho\left (Y,X_1,X_2,\tau\right ),
\end{flalign}
\end{subequations}
where the specific form of $\xi\left (Y,X_1,X_2,\tau\right )$, $\eta\left
  (Y,X_1,X_2,\tau\right )$, and $\rho\left (Y,X_1,X_2,\tau\right )$ is chosen
to simplify the original system according to the five principles listed
previously, and is found using an iterative procedure.  To outline the
procedure, we provide details for a simple example in Appendix~\ref{sec:simple_example}.

Several iterations lead to coordinate transforms for $U$, $V$, and $W$ along
with evolution equations describing the $Y$-dynamics, $X_1$-dynamics, and
$X_2$-dynamics in the new coordinate system. The $Y$-dynamics have exponential
decay to the $Y=0$ slow manifold. Substitution of $Y=0$ leads to the coordinate transforms
\begin{subequations}
\begin{flalign}
U =& 
 \sigma_4 {\cal G} \left(  \phi_4 \right) 
+{\frac {{ \gamma_0}^2 \left( \sigma_6 {\cal G} \left(  \phi_6 \right) - X_2 \right) }{ \left(  \alpha_0+ \gamma_0 \right)^2}}
+\mu  \left[ {\frac { \gamma_0 \beta  \left[  \sigma_5  X_2 {\cal G} \left(  \phi_5 \right) - X_1  X_2+ \sigma_6  X_1 {\cal G} \left(  \phi_6 \right)  \right] }{ \left(  \alpha_0+ \gamma_0 \right)^2}} - \right. &&\nonumber \\
& \left. \frac{\sigma_4  \alpha_0 \beta  \gamma_0 {\cal G}^2 \left(  \phi_4 \right)  \left[ (\alpha_0+ \gamma_0) X_1+\gamma_0  X_2 \right]}{(\alpha_0+ \gamma_0)\left(  \gamma_0+\mu^2 \right) \left(  \alpha_0+\mu^2 \right)}    \right] 
+ \mu^2 \left[ {\frac { \gamma_0  \left[  X_1  \alpha_0-2  X_2+2  \sigma_6 {\cal G} \left(  \phi_6 \right)  \right] }{ \alpha_0  \left(  \alpha_0+ \gamma_0 \right) }} - \right. \nonumber \\
& \left. \frac{\sigma_4 {\cal G}^2 \left(  \phi_4 \right)  }{ \left(  \gamma_0+\mu^2 \right) \left(  \alpha_0+\mu^2 \right)}  \left( {\frac {2 { \gamma_0}^{3}+{ \alpha_0}^{3}}{ \alpha_0+ \gamma_0}}+{\frac {{ \alpha_0}^2{ \gamma_0}^2 \left(  \alpha_0+ \gamma_0 \right) }{ \left(  \gamma_0+\mu^2 \right)  \left(  \alpha_0+\mu^2 \right) }} \right)  
-{\frac { \sigma_5  \gamma_0 {\cal G} \left(  \phi_5 \right) }{ \left(  \alpha_0+ \gamma_0 \right)^2}} \right] + \nonumber \\
& \mu^{3} \left[ {\frac {\beta  \left[ - X_1+ \sigma_5 {\cal G} \left(  \phi_5 \right)  \right] }{ \left(  \alpha_0+ \gamma_0 \right)^2}}- \frac{\sigma_4 \beta {\cal G}^2 \left(  \phi_4 \right)  }{ \left(  \gamma_0+\mu^2 \right) \left(  \alpha_0+\mu^2 \right)} \left( {\frac { \alpha_0  \gamma_0}{ \alpha_0+ \gamma_0}}+ \gamma_0  X_2+ X_1  \left(  \alpha_0+ \gamma_0 \right)  \right) + \right. \nonumber \\
& \left. {\frac {\beta  \left(  \sigma_6  X_1 {\cal G} \left(  \phi_6 \right) + \sigma_5  X_2 {\cal G} \left(  \phi_5 \right) - X_1  X_2  \alpha_0 \right) }{ \alpha_0  \left(  \alpha_0+ \gamma_0 \right) }} \right] +\mathcal{O}(\mu^4)
,\label{e:U_ct_mort}
\end{flalign}
\begin{flalign}
V =& X_1 +\mu \left [\frac{\sigma_4\alpha_0\beta X_1 {\cal G}(\phi_4)}{\alpha_0+\gamma_0}+\frac{\sigma_4\alpha_0\beta X_2{\cal G}(\phi_4)}{\left (\alpha_0+\gamma_0\right
  )^2} \right ]+ \mu^2\left [\frac{\sigma_4 {\cal G}(\phi_4) \left (
   \alpha_0^2+\alpha_0\gamma_0+\gamma_0^2\right )}{\gamma_0\left
   (\alpha_0+\gamma_0\right )^2} \right ] + &&\nonumber\\
&\mu^3\left [\frac{\sigma_4\alpha_0\beta {\cal G}(\phi_4)}{\gamma_0\left (\alpha_0+\gamma_0 \right )^2}+\frac{\sigma_4\beta X_2{\cal G}(\phi_4)}{\gamma_0\left
   (\alpha_0+\gamma_0\right )} 
 +\frac{\sigma_4\beta X_1{\cal G}(\phi_4) }{\gamma_0^2} \right
]+\mathcal{O}(\mu^4),\label{e:V_ct_mort}
\end{flalign}
\begin{flalign}
W =& X_2 +\mu\left
 [-\frac{\sigma_4\beta X_1{\cal G}(\phi_4)}{\gamma_0}
 -\frac{\sigma_4\beta X_2{\cal G}(\phi_4)}{\left (\alpha_0+\gamma_0\right
  )} \right ]+ \mu^2\left [\frac{\sigma_4 {\cal G}(\phi_4) \left (
   \alpha_0^2+\alpha_0\gamma_0+\gamma_0^2\right )}{\alpha_0\gamma_0\left
   (\alpha_0+\gamma_0\right )} \right ]+ &&\nonumber\\
&\mu^3\left [-\frac{\sigma_4\beta {\cal G}(\phi_4) }{\gamma_0\left (\alpha_0+\gamma_0\right
  )}-\frac{\sigma_4 \left (\alpha_0+\gamma_0\right
  )\beta X_1{\cal G}(\phi_4) }{\alpha_0\gamma_0^2} - \frac{\sigma_4\beta X_2{\cal G}(\phi_4)}{\alpha_0\gamma_0}\right ]+\mathcal{O}(\mu^4),\label{e:W_ct_mort}
\end{flalign}
\end{subequations}
where 
\begin{equation}
\label{e:conv}
{\cal G}(\phi) = e^{-\aleph\tau}*\phi = \int\limits_{-\infty}^{\tau}
\exp{[-\aleph\cdot (\tau -s)]\phi(s)} ds, ~~~
\aleph = \frac{\alpha_0\gamma_0\left (\alpha_0 +\gamma_0\right )}{\left
  (\alpha_0+\mu^2\right )\left (\gamma_0+\mu^2\right )},
\end{equation}
and 
\begin{equation}
\label{e:conv2}
{\cal G}^2(\phi) = e^{-\aleph\tau}*e^{-\aleph\tau}*\phi.
\end{equation}

All of the stochastic terms in Eqs.~(\ref{e:U_ct_mort})-(\ref{e:W_ct_mort})
consist of integrals of the noise process into the past (convolutions), as
given by Eqs.~(\ref{e:conv}) and~(\ref{e:conv2}).  These memory integrals are
fast-time processes.  Since we are interested in the long-term slow processes
and since the expectation of ${\cal G}$ equals $e^{-\aleph\tau}*E[\phi]$,
where $E[\phi]=0$, we neglect the memory integrals and the higher-order
multiplicative terms found in Eqs.~(\ref{e:U_ct_mort})-(\ref{e:W_ct_mort})
so that
\begin{subequations}
\begin{flalign}
U = &  - \frac{\gamma_0^2 X_2}{\left (\alpha_0 + \gamma_0\right )^2}
-  \frac{\mu\beta X_1}{\left (\alpha_0 + \gamma_0\right )} \left(
\frac{\mu^2}{\left (\alpha_0 + \gamma_0\right )}
+\frac{\gamma_0 X_2}{\left (\alpha_0 + \gamma_0\right )}
+ \mu^2 X_2 \right)&& \nonumber\\
&+\frac{\mu^2\gamma_0}{\left (\alpha_0+\gamma_0\right )}\left( X_1-\frac{2X_2}{\alpha_0} \right)
,\label{e:U_cm_mort}\\
V = & ~ X_1,&&\\
W = & ~ X_2.&&
\end{flalign}
\end{subequations}
Note that Eq.~(\ref{e:U_cm_mort}) is the deterministic center manifold
equation, and at first-order, matches the center manifold equation that was
found previously [Eq.~(\ref{e:CMeq})].

\begin{figure}[t!]
\includegraphics[width=8.5cm,height=5cm]{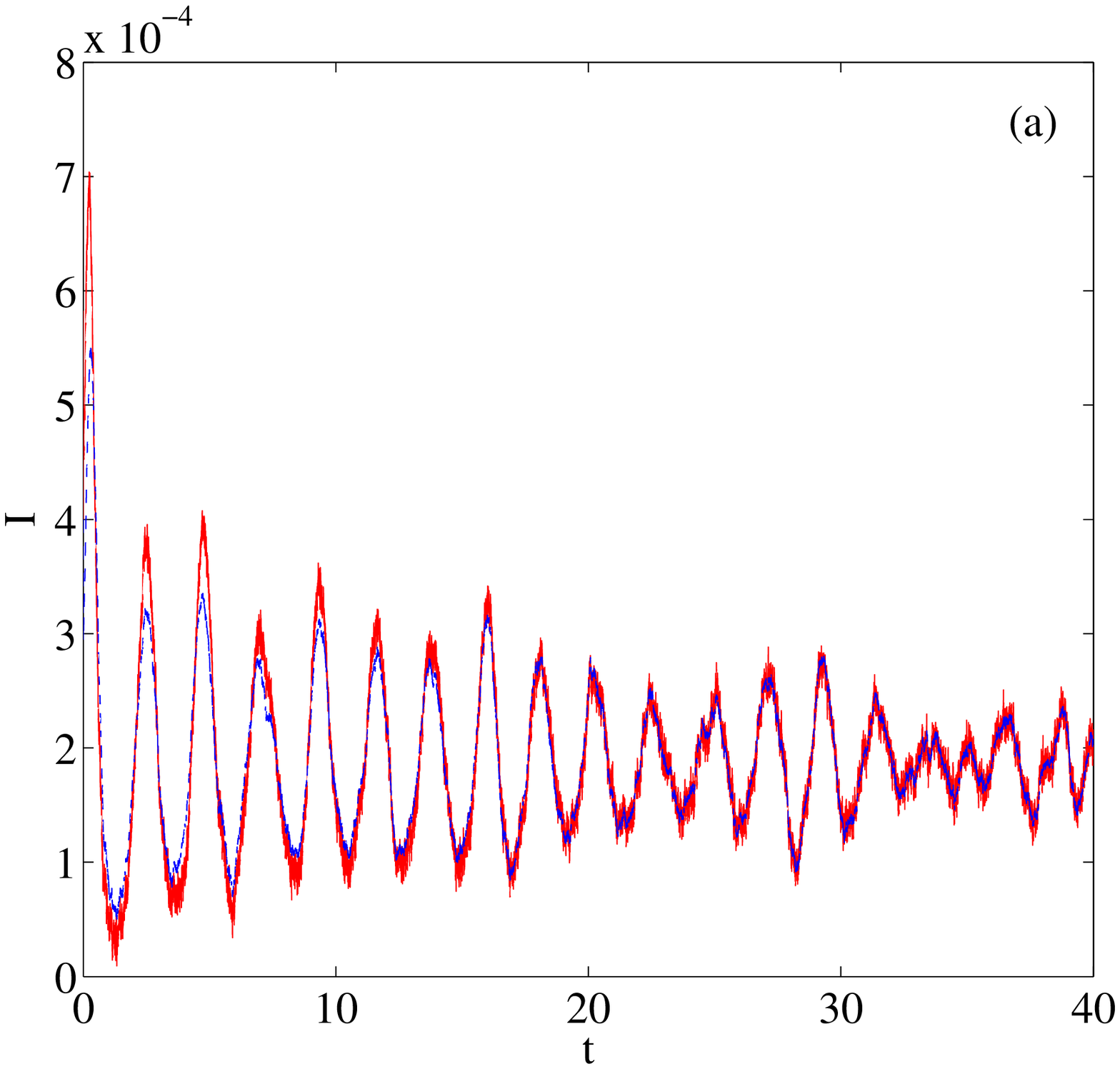}\\
\includegraphics[width=8.5cm,height=5cm]{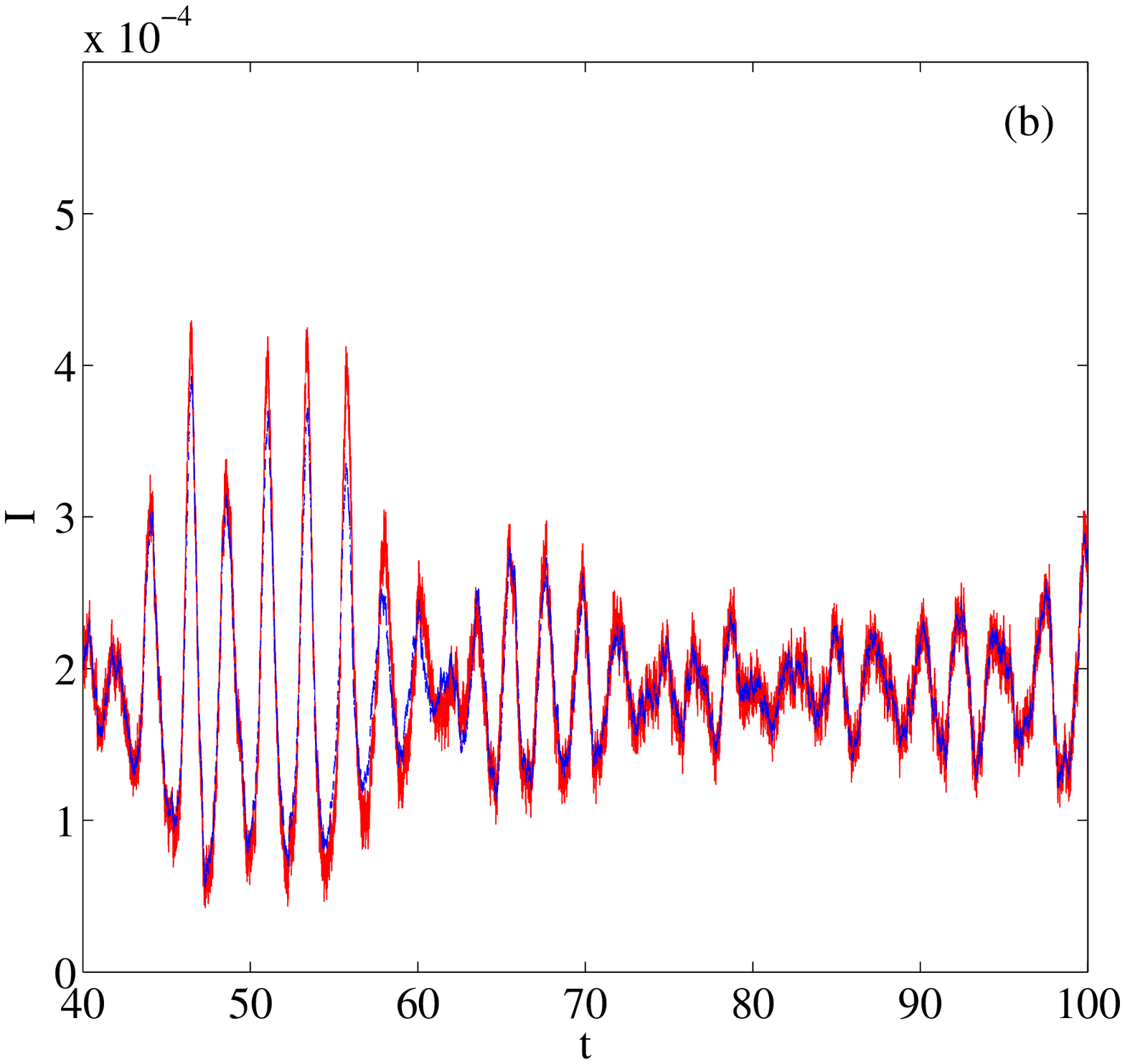}
\caption{\label{fig:TS_full_reducedNF}(Color online) Time series of the
  fraction of the population that is  infected with a disease, $I$, computed
  using the complete, stochastic system of transformed equations of the SEIR
  model [Eqs.~(\ref{e:dU_mort_stoch})-(\ref{e:dW_mort_stoch})] (red, solid
  line), and computed using the reduced system of equations of the SEIR model
  that is found using the stochastic normal form coordinate transform
  [Eqs.~(\ref{e:F}), (\ref{e:G}), (\ref{e:x1_NFCT_mort}),
  and~(\ref{e:x2_NFCT_mort})] (blue, dashed line). The standard deviation of
  the noise intensity used in the simulation is $\sigma_i =0.0005$,
  $i=4,5,6$.  The time series is shown for (a)
  $t=0$ to $t=40$, and for (b) $t=40$ to $t=100$.}
\end{figure}
Substitution of $Y=0$ and neglecting all multiplicative noise terms and memory
integrals using the argument from above (so that we consider only first-order noise terms) leads to the
following reduced system of evolution equations on the center manifold:
\begin{subequations}
\begin{flalign}
\frac{dX_1}{d\tau} = F(X_1(\tau),X_2(\tau)),\label{e:F}\\
\frac{dX_2}{d\tau} = G(X_1(\tau),X_2(\tau))\label{e:G}.
\end{flalign}
\end{subequations}
The specific form of $F$ and $G$ in Eqs.~(\ref{e:F}) and~(\ref{e:G}) are complicated, and are
therefore presented in Appendix~\ref{sec:cpcm_mort}.

We numerically integrate the complete, stochastic system of transformed
equations of the SEIR model [Eqs.~(\ref{e:dU_mort_stoch})-(\ref{e:dW_mort_stoch})] along with the reduced
system of equations that is found using the stochastic normal form coordinate transform
[Eqs.~(\ref{e:F}), (\ref{e:G}), (\ref{e:x1_NFCT_mort}), and~(\ref{e:x2_NFCT_mort})]. The complete system
is solved for $U$, $V$, and $W$, while the reduced system is solved for
$X_1=V$ and $X_2=W$. In the latter case, $U$ is computed using the center
manifold equation given by Eq.~(\ref{e:U_cm_mort}). As before, once the values of $U$, $V$, and $W$ are known, we compute the values of $\bar{S}$, $\bar{E}$ , and $\bar{I}$ using the transformation given by Eqs.~(\ref{e:U})-(\ref{e:W}). We shift $\bar{S}$, $\bar{E}$, and $\bar{I}$ respectively by $S_0$, $E_0$, and $I_0$ to find the values of $S$, $E$, and $I$.

Figures~\ref{fig:TS_full_reducedNF}(a)-(b) compares the time series of the fraction of
the population that is infected with a disease, $I$, computed using the
complete, stochastic system of transformed equations of the SEIR model
[Eqs.~(\ref{e:dU_mort_stoch})-(\ref{e:dW_mort_stoch})] with the time series of
$I$ computed using the reduced system of equations of the SEIR model that is found using
the stochastic normal form coordinate transform [Eqs.~(\ref{e:F}),
(\ref{e:G}), (\ref{e:x1_NFCT_mort}), and~(\ref{e:x2_NFCT_mort})].
Figure~\ref{fig:TS_full_reducedNF}(a) shows the initial transients, while
Fig.~\ref{fig:TS_full_reducedNF}(b) shows the time series after the transients
have decayed.  One can see that there is excellent agreement between the two
solutions.  The initial outbreak is successfully captured by the reduced
system.  Furthermore, Fig.~\ref{fig:TS_full_reducedNF}(b) shows that the reduced system accurately predicts recurrent
outbreaks for a time scale that is orders of magnitude longer than the
relaxation time.  This is not surprising since the solution decays
exponentially throughout the transient and then remains close to the
lower-dimensional center manifold.  Since we are not looking at periodic
orbits, there are no secular terms in the asymptotic expansion, and the result
is valid for all time.  Additionally, any noise drift on the center manifold
results in bounded solutions due to sufficient dissipation transverse to the manifold.  The cross-correlation of the two time series shown in
Fig.~\ref{fig:TS_full_reducedNF} is approximately 0.98, which quantitatively suggests
there is excellent agreement between the two solutions.

\begin{figure}[h!]
\includegraphics[width=8.5cm,height=7cm]{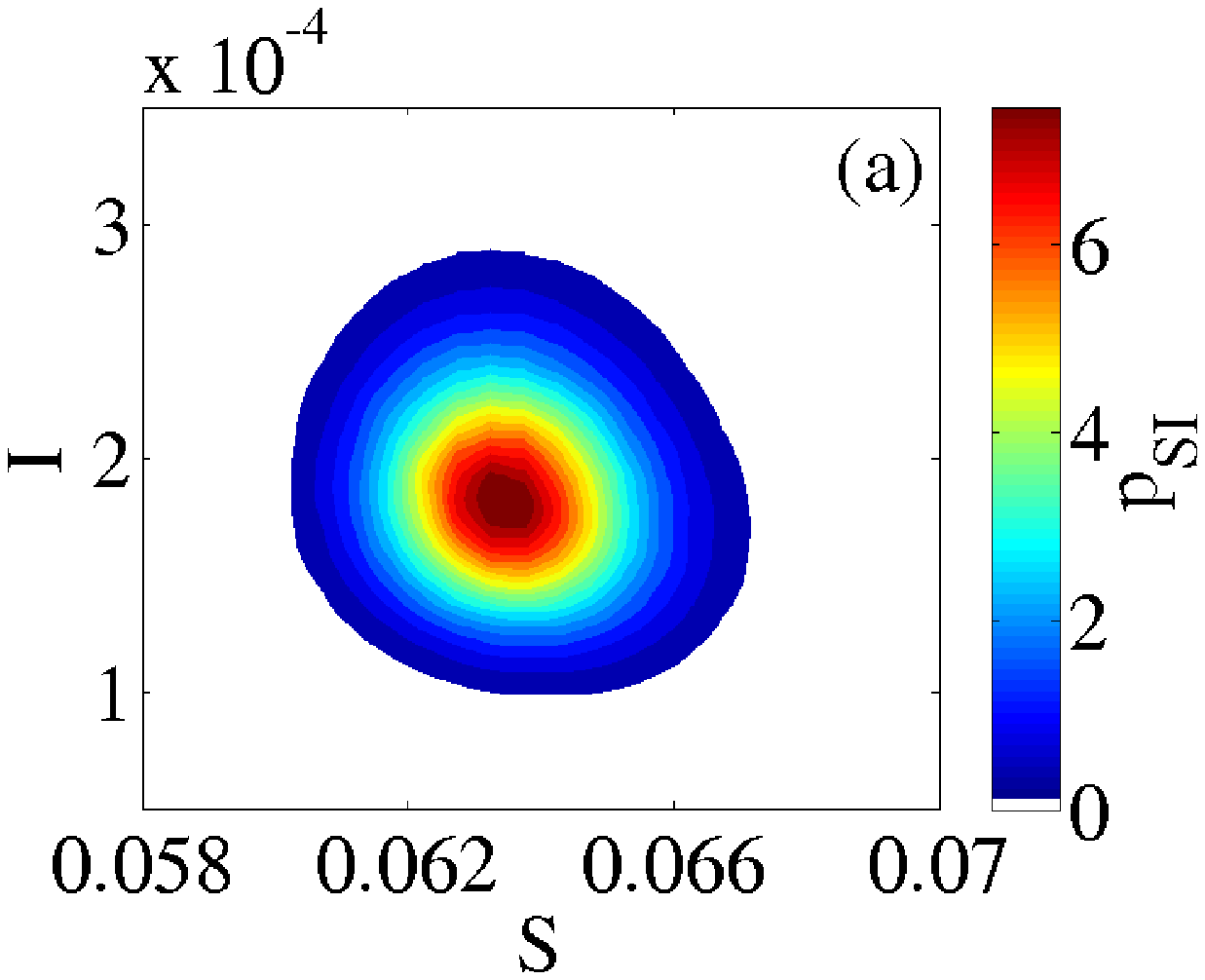}
\includegraphics[width=8.5cm,height=7cm]{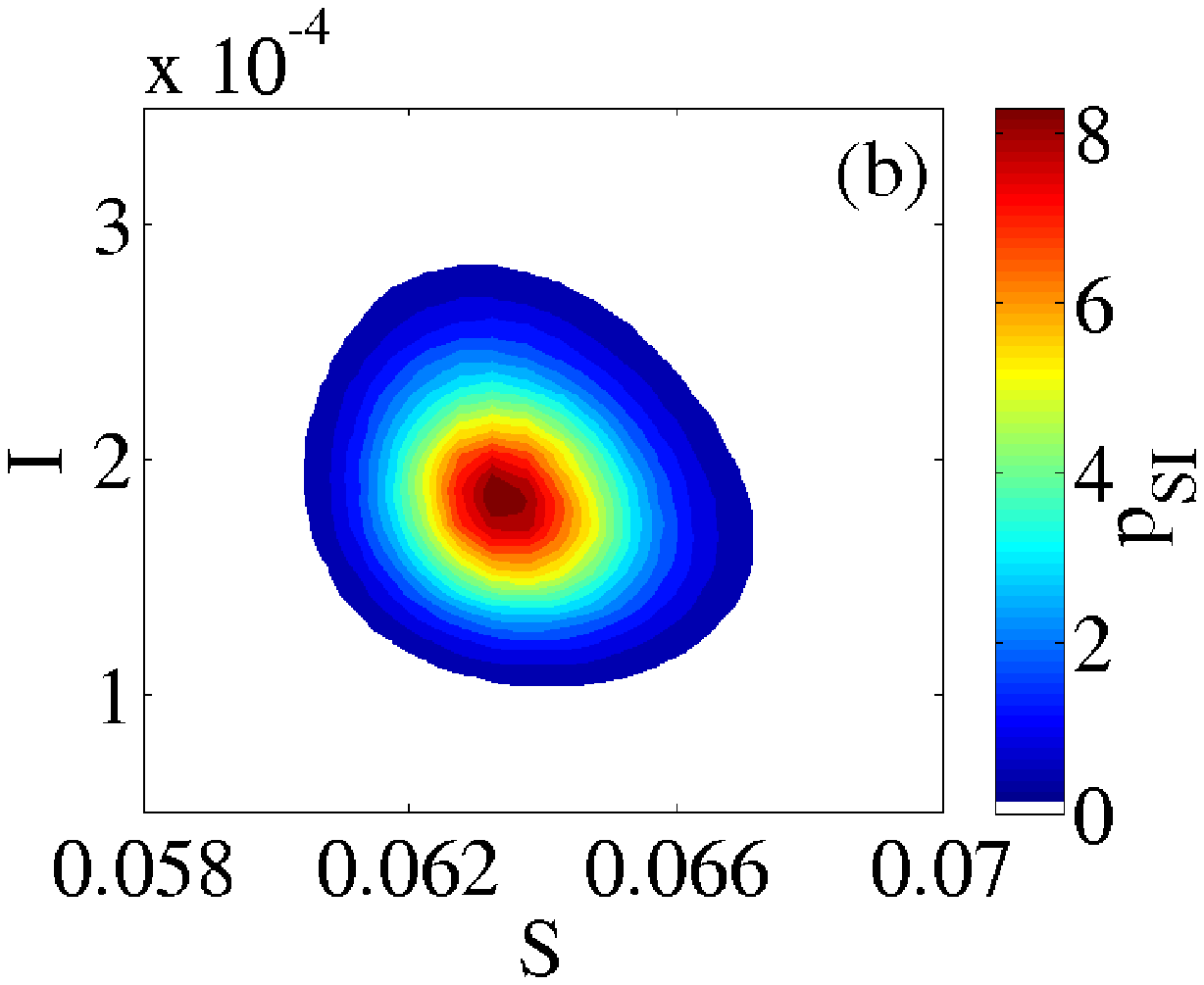}
\caption{\label{fig:hist_full_reducedNF}(Color online) Histogram of
  probability density, $p_{SI}$ of the $S$ and $I$ values found using (a) the
  complete, stochastic  system of transformed equations for the SEIR model
  with mortality [Eqs.~(\ref{e:dU_mort_stoch})-(\ref{e:dW_mort_stoch})], and
  (b)  the reduced system of equations of the SEIR model with mortality that is found using the  stochastic normal form coordinate transform [Eqs.~(\ref{e:F}), (\ref{e:G}), (\ref{e:x1_NFCT_mort}), and~(\ref{e:x2_NFCT_mort})]. The  histograms are created using $100$ years worth of time series (starting with year $40$) for 500 realizations, and the color-bar values have been normalized by $10^{-3}$.}
\end{figure}

Using the same systems of transformed equations, we compute $140$ years worth
of time series for $500$ realizations. As before, we ignore the first $40$
years worth of transient solution, and the data is used to create a histogram
representing the probability density, $p_{SI}$ of the $S$ and $I$
values. Figure~\ref{fig:hist_full_reducedNF}(a) shows the histogram associated
with the complete, stochastic system of transformed equations, while
Fig.~\ref{fig:hist_full_reducedNF}(b) shows the histogram associated with the
reduced system of equations found using the normal form coordinate transform. The color-bar values in Figs.~\ref{fig:hist_full_reducedNF}(a)-(b) have been normalized by $10^{-3}$.

As we saw with Figs.~\ref{fig:hist_full_reducedCM}(a)-(b), the probability
distribution shown in Fig.~\ref{fig:hist_full_reducedNF}(a) looks
qualitatively the same as the probability distribution shown in
Fig.~\ref{fig:hist_full_reducedNF}(b). Using the Kullback-Leibler divergence
given by Eq.~(\ref{e:KL}), we have found that the two histograms shown in
Figs.~\ref{fig:hist_full_reducedNF}(a)-(b) have a relative entropy of
$d_{KL}=0.0953$. Since this value is close to zero, the two histograms are
quantitatively very similar.

\begin{figure}[t!]
\includegraphics[width=8.5cm,height=6.375cm]{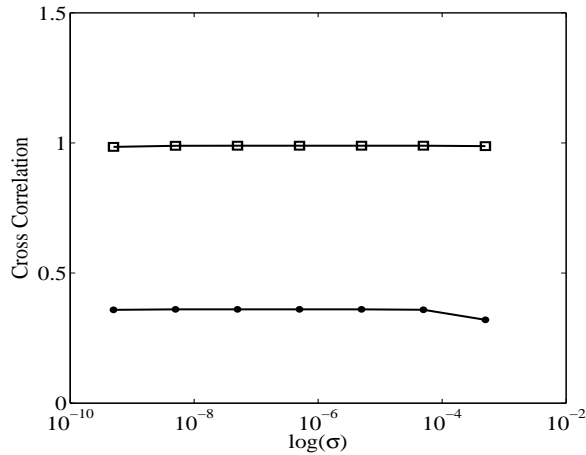}
\caption{\label{fig:CC} Cross-correlation between time series found using the
  original, stochastic system of transformed equations and the reduced system
  of equations based on the deterministic center manifold (``circle'' markers),
  and cross-correlation between time series found using the
  original, stochastic system of transformed equations and the reduced system
  of equations based on the stochastic normal form coordinate transform
  (``square'' markers).  The cross-correlation is computed using time series from $t=800$ to $t=1000$.}
\end{figure}

In addition to computing the cross-correlation between the solution of the
original system and the solutions of the two reduced systems for $\sigma_i
=0.0005$, we have computed the cross-correlation for the case of zero noise as
well as for noise intensities ranging from $\sigma =5.0\times 10^{-10}$ to
$\sigma =5.0\times
10^{-5}$.  These cross-correlations were computed using time series from
$t=800$ to $t=1000$.  For the deterministic case (zero noise), the cross-correlation
between the time series which were computed using the original system and the reduced
system based on the deterministic center manifold is $1.0$, since the
agreement is perfect.  The cross-correlation between the original system and
the reduced system found using the stochastic normal form is also $1.0$.
Figure~\ref{fig:CC} shows the cross-correlation between the original system
and the two reduced systems for various values of $\sigma$.

One can see in Fig.~\ref{fig:CC} that the solutions found using the reduced system
based on the deterministic center manifold compare poorly with the
original system at very low noise values.  Furthermore, as the noise
increases, the agreement between the two solutions gets worse.  On the other
hand, Fig.~\ref{fig:CC} shows that the solutions computed using the reduced
system found using the normal form
coordinate transform compare very well with the solutions to the original
system across a wide range of small noise intensities.

\section{Discussion}\label{sec:disc}
We have demonstrated that the normal form coordinate transform method
  reduces the Langevin system so that both the noise and dynamics are
  accurately projected onto the lower-dimensional manifold.  It is natural to
  consider (a) the replacement of the stochastic term by a deterministic, period
  drive of small amplitude, and (b) the extension to finite populations.
  These cases are discussed respectively in Sec.~\ref{sec:DF} and Sec.~\ref{sec:FP}.
\subsection{The Case of Deterministic Forcing}\label{sec:DF}
A single time series realization of the noise might be thought of as a
deterministic function of small amplitude driving the system. One could
rederive the normal form for such a deterministic function. However,
since our derived  normal form holds specifically for the case of white
noise,  we show that a simple replacement of the stochastic realization with a
deterministic realization does not work.  As an example, one
could consider the following sinusoidal functions:
\begin{subequations}
\begin{flalign}
&\sigma_1\phi_1 = \cos{(10\pi\mu t)}/8000,\label{e:u1}\\
&\sigma_2\phi_2 = \sin{(4\pi\mu t)}/8000,\label{e:u2}\\
&\sigma_3\phi_3 = \cos{(10\pi\mu t)}/8000,\label{e:u3}
\end{flalign}
\end{subequations}
where $\sigma_4\phi_4$, $\sigma_5\phi_5$, and
$\sigma_6\phi_6$ are given by Eqs.~(\ref{e:sig4phi4})-(\ref{e:sig6phi6}).
Using Eqs.~(\ref{e:u1})-(\ref{e:u3}) or some other similar deterministic
drive, the solution computed using the reduced system based on the
deterministic center manifold analysis will agree perfectly with the solution
computed using the complete system of equations.  On the other hand, since the
reduced system based on the normal form analysis was derived specifically for
white noise, the transient solution found using this reduced system will not
agree with the solution found using the complete system.  It is possible to
find a normal form coordinate transform for periodic forcing, but the normal
form will be different than the one derived in this article for white noise.

\begin{figure}[t!]
\includegraphics[width=8.5cm,height=5cm]{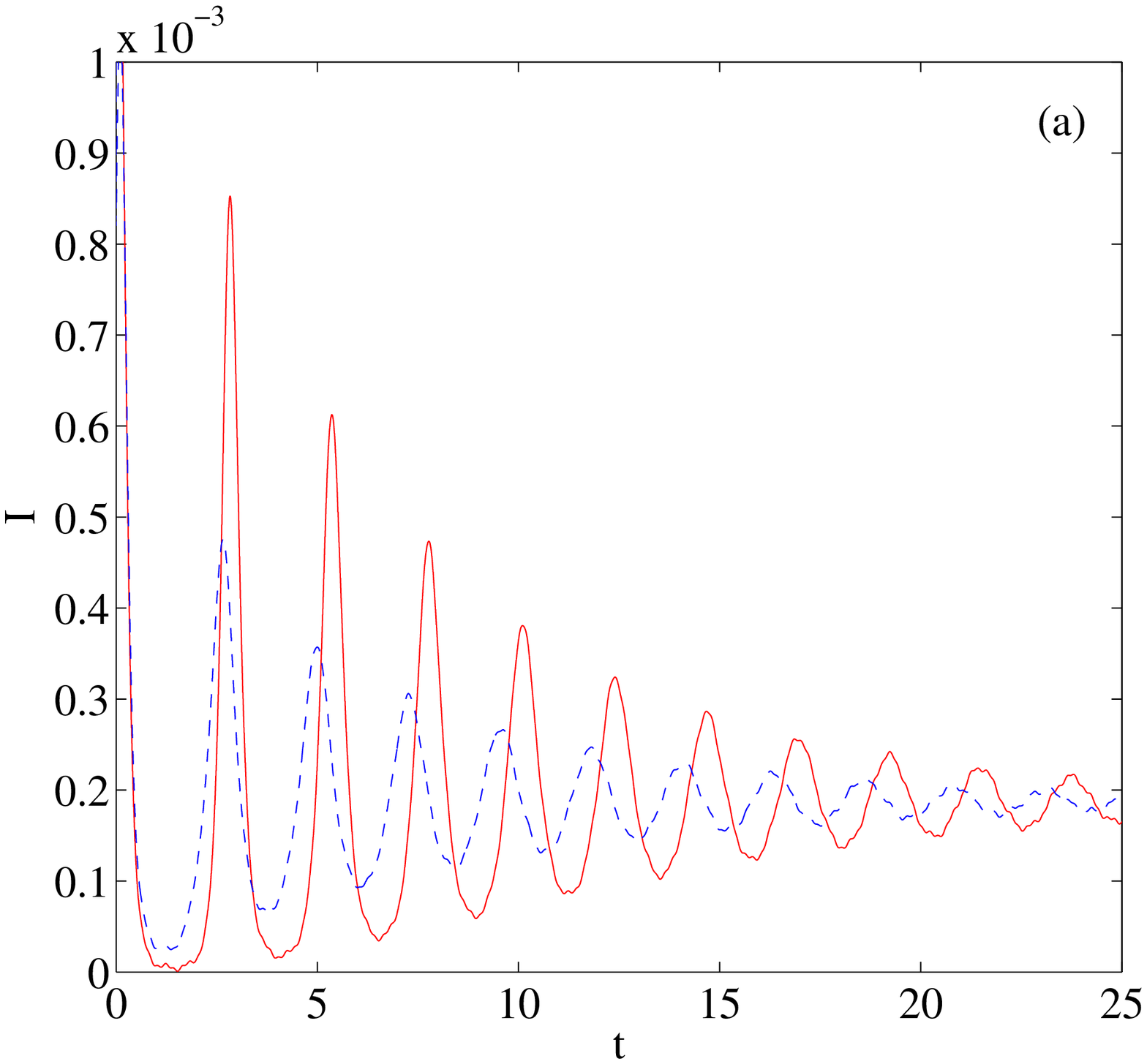}\\
\includegraphics[width=8.5cm,height=5cm]{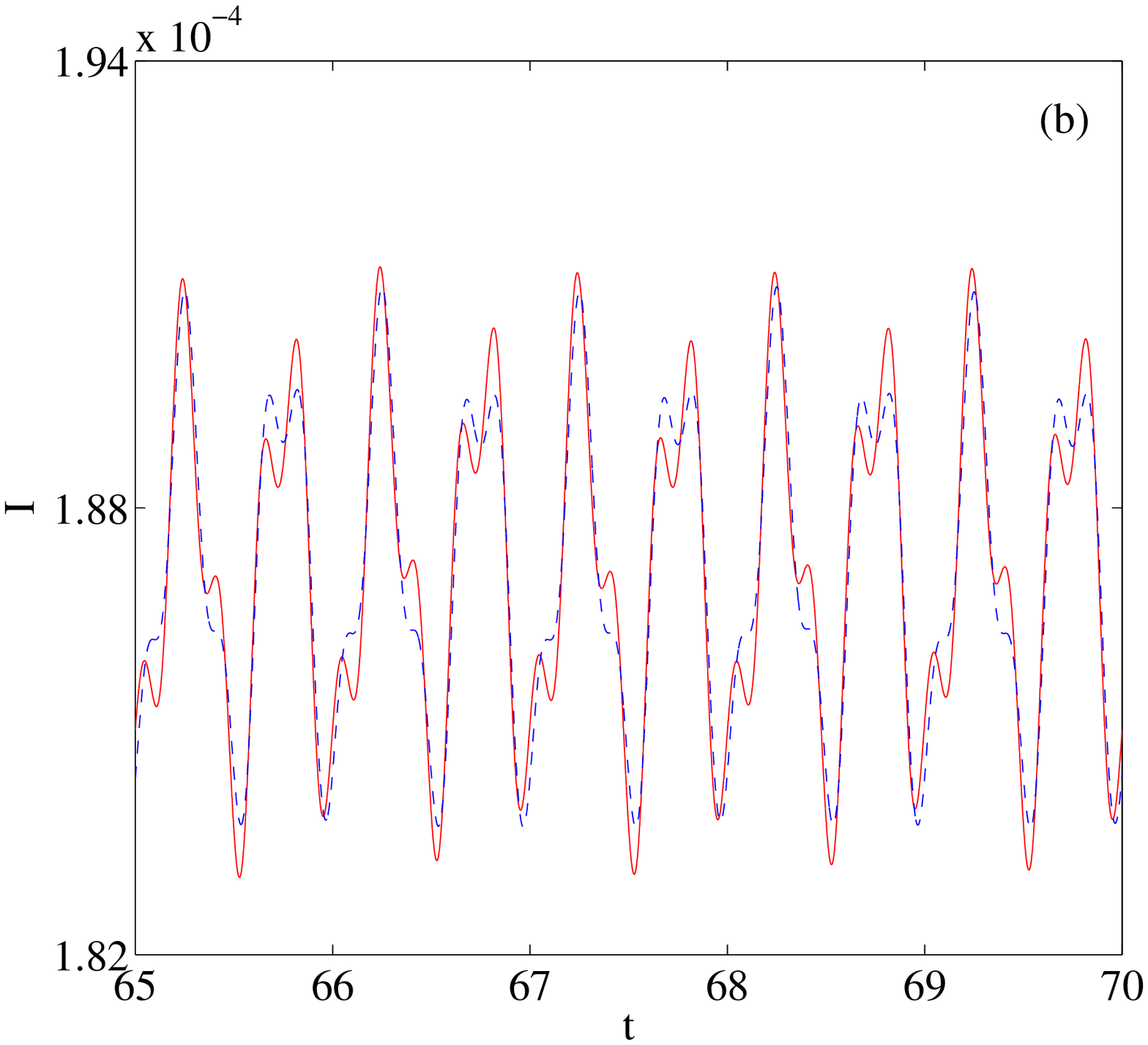}
\caption{\label{fig:determ}(Color online) Time series of the
  fraction of the population that is  infected with a disease, $I$, computed
  using the complete system of transformed equations of the SEIR
  model [Eqs.~(\ref{e:dU_mort_stoch})-(\ref{e:dW_mort_stoch})] (red, solid
  line), and computed using the reduced system of equations of the SEIR model
  that is found using the normal form coordinate transform [Eqs.~(\ref{e:F}),
  (\ref{e:G}), (\ref{e:x1_NFCT_mort}), and~(\ref{e:x2_NFCT_mort})] (blue,
  dashed line). The stochastic terms in both systems have been replaced by the
deterministic terms given by Eqs.~(\ref{e:u1})-(\ref{e:u2}).  The
time series is shown from (a) $t=0$ to $t=25$, and from (b) $t=65$ to
$t=70$. } 
\end{figure}

Figures~\ref{fig:determ}(a)-(b) compares the time series of the fraction of
the population that is infected with a disease, $I$, computed using the
complete system of transformed equations of the SEIR model [Eqs.~(\ref{e:dU_mort_stoch})-(\ref{e:dW_mort_stoch})] with the time series of
$I$ computed using the reduced system of equations of the SEIR model that is found using
the stochastic normal form coordinate transform [Eqs.~(\ref{e:F}),
(\ref{e:G}), (\ref{e:x1_NFCT_mort}), and~(\ref{e:x2_NFCT_mort})], but where
the stochastic terms of both systems have been replaced by the deterministic
terms given by Eqs.~(\ref{e:u1})-(\ref{e:u3}).  Figure~\ref{fig:determ}(a) shows the initial transients, while
Fig.~\ref{fig:determ}(b) shows a piece of the time series after the transients
have decayed.  One can see in Figs.~\ref{fig:determ}(a)-(b) that although the
two solutions eventually become relatively synchronized with one another,
there is poor agreement, both in phase and amplitude, throughout the transient.

\subsection{The Case of Finite Populations}\label{sec:FP}
The solutions to the original system and both reduced systems are continuous
solutions based on an infinite population assumption, and are found using Langevin
equations having Gaussian noise.  It is interesting to examine the effects of
general noise by using a Markov simulation to compare solutions of the
original and reduced systems.

\begin{figure}[h!]
\includegraphics[width=8.5cm,height=6cm]{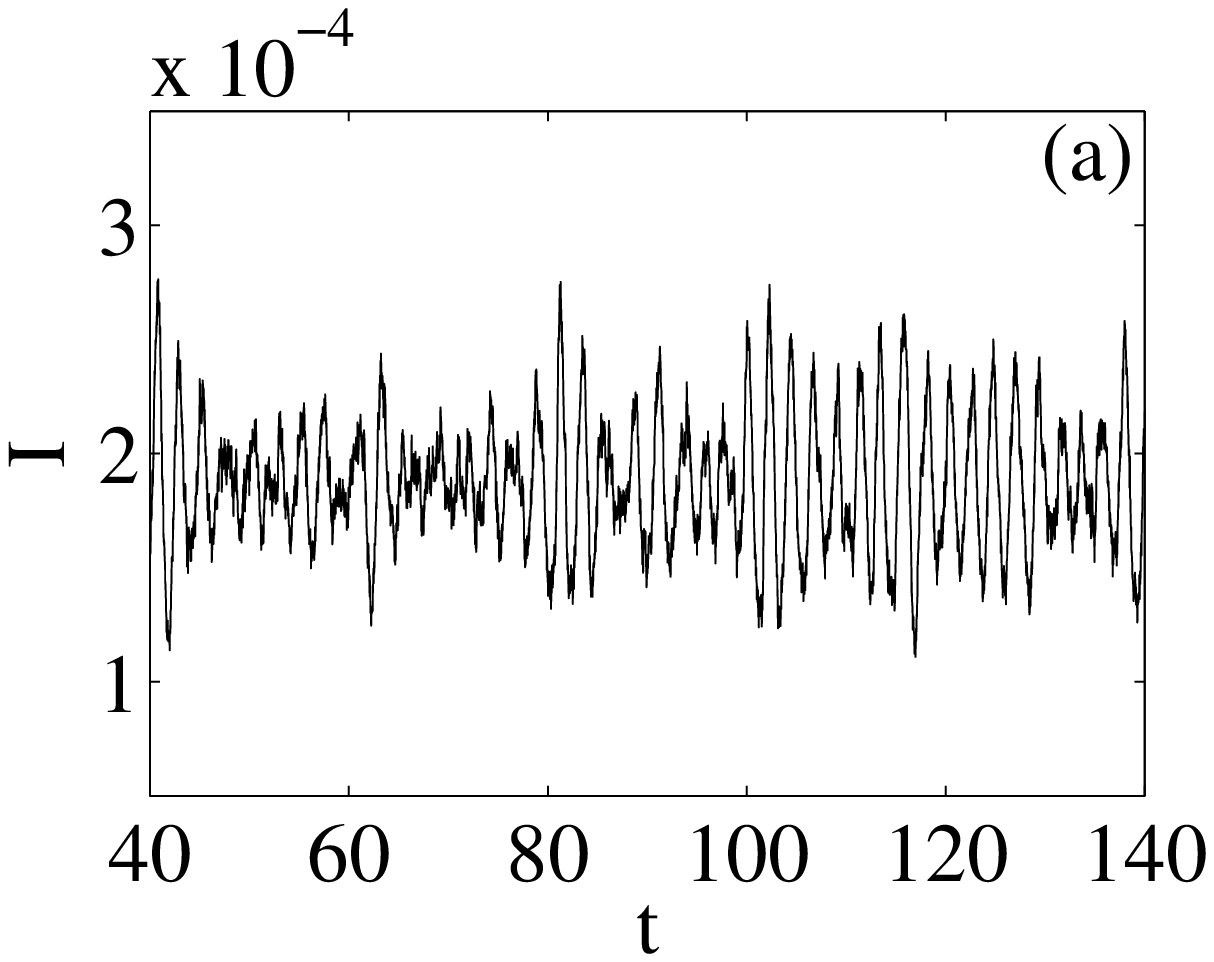}
\includegraphics[width=8.5cm,height=7cm]{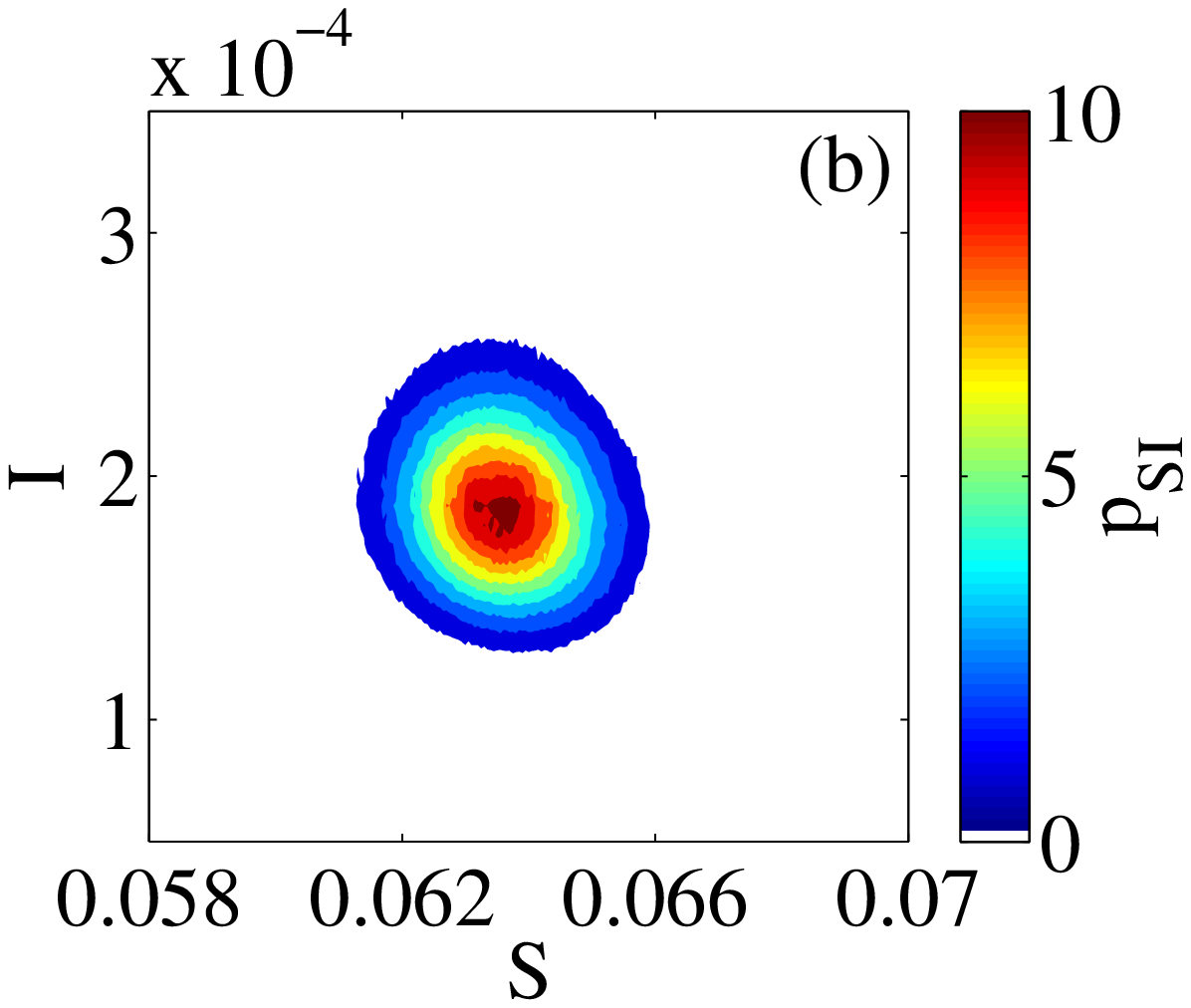}
\caption{\label{fig:simulationSEI} (a) Time series of the fraction of the
  population that is infected with a disease, $I$, computed using a Markov
  simulation of the complete, original equations of the SEIR model
  [Eq.~(\ref{e:evolution})], and (b) (color online) a histogram of probability density,
  $p_{SI}$ of the $S$ and $I$ values found using a Markov
  simulation of Eq.~(\ref{e:evolution}).  The histogram is created using $100$
  years worth of data (starting with year $40$) for 500 realizations, and the
  color-bar values have been normalized by $10^{-4}$.}
\end{figure}

The complete system in the original variables (see page 5) will evolve in time $t$ in the following way: 
\begin{equation} \label{e:evolution}
\begin{array}{lcl}
\mbox{transition}  &\hspace*{0.35in}  & \mbox{rate}  \\
(s-1,e+1,i)  &&  \beta s i /N        \\
(s,e-1,i+1)  &&  \alpha e            \\
(s,e,i-1)    &&  \gamma i            \\
(s+1,e,i)  &&  \mu N                  \\
(s-1,e,i)    &&  \mu s               \\
(s,e-1,i)  &&  \mu e               \\
(s,e,i-1)  &&  \mu i               
\end{array} .
\end{equation}
Using a total population size of $N=10$ million, we have performed a Markov simulation of the
system.  After completing the Markov simulation, we divided $s$, $e$, and $i$
by $N$ to find $S$, $E$, and $I$.  Figure~\ref{fig:simulationSEI}(a) shows a time series, after the
transients have decayed, of the fraction
of the population that is infected with a disease, $I$.  The results reflect
both the mean
and the frequency of the deterministic system.  Performing the simulation for
$500$ realizations allows us to create a histogram representing the
probability density, $p_{SI}$ of the $S$ and $I$ values.  This histogram is
shown in Fig.~\ref{fig:simulationSEI}(b), and one can see that the probability
density reflects the
amplitude, which varies with the population size, of $S$ and $I$.  The color-bar values in
Fig.~\ref{fig:simulationSEI}(b) have been normalized by $10^{-4}$ 

The complete system in the transformed variables has the stable endemic
equilibrium at the origin. To bound the dynamics to the first octant, we use
the fact that $s \ge 0$, $e \ge 0$, and $i \ge 0$ to derive the appropriate
inequalities for the transformed, discrete variables $u$, $v$, and $w$.  These
inequalities can be found in Appendix~\ref{sec:MSim} as Eq.~(\ref{e:ineq}).
These inequalities enable us to define new discrete variables $Y_1$, $Y_2$, and $Y_3$,
given by Eqs.~(\ref{e:Y1})-(\ref{e:Y3}) in Appendix~\ref{sec:MSim}. 

In the $Y_i$ variables, we define evolution relationships similar to those
found in Eq.~(\ref{e:evolution}). The complete transformed system will evolve in time
$\tau$ according to the transition and rates given by Eq.~(\ref{e:Vevolution})
in Appendix~\ref{sec:MSim}.  

\begin{figure}[!h]
\includegraphics[width=8.5cm,height=6cm]{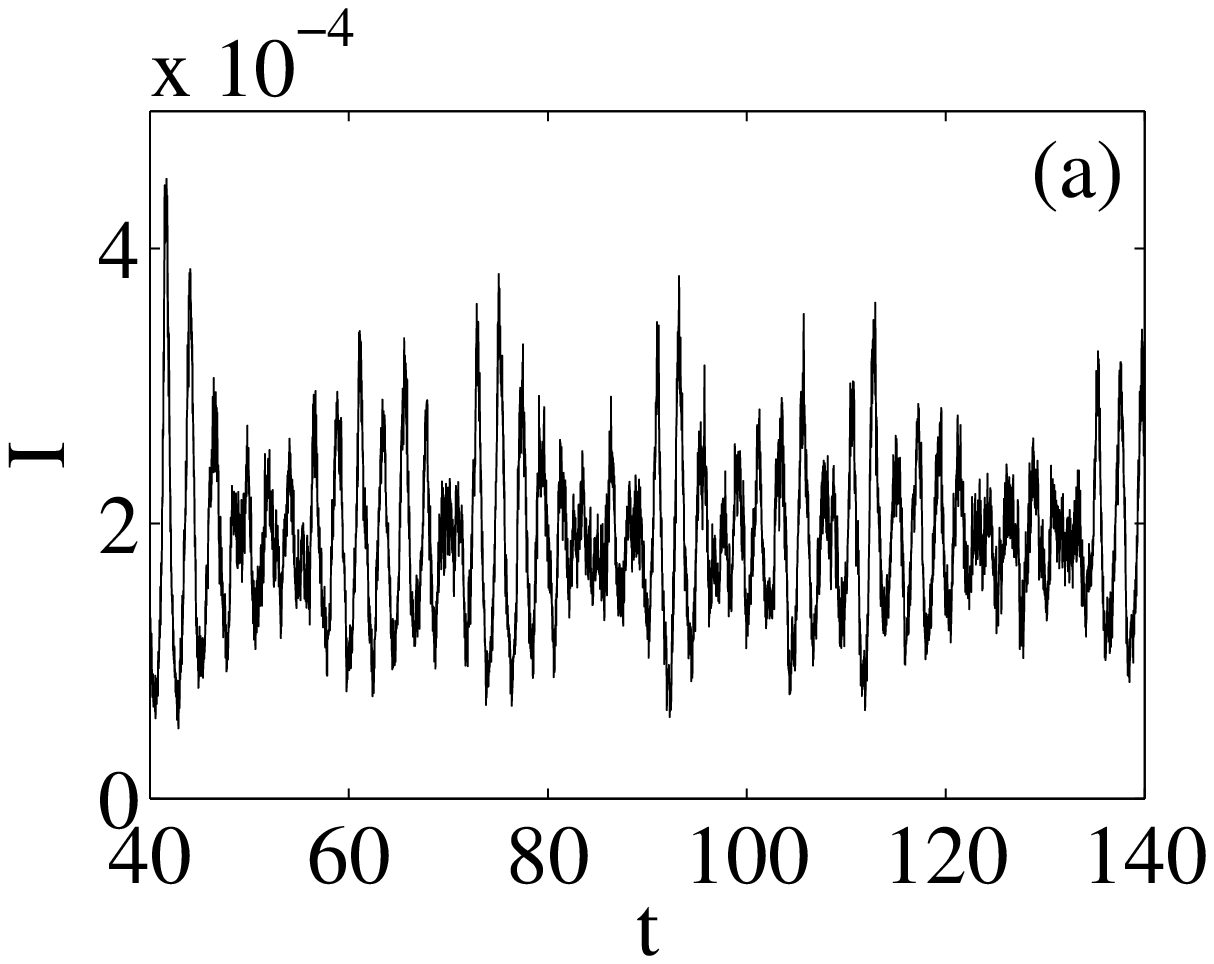}
\includegraphics[width=8.5cm,height=7cm]{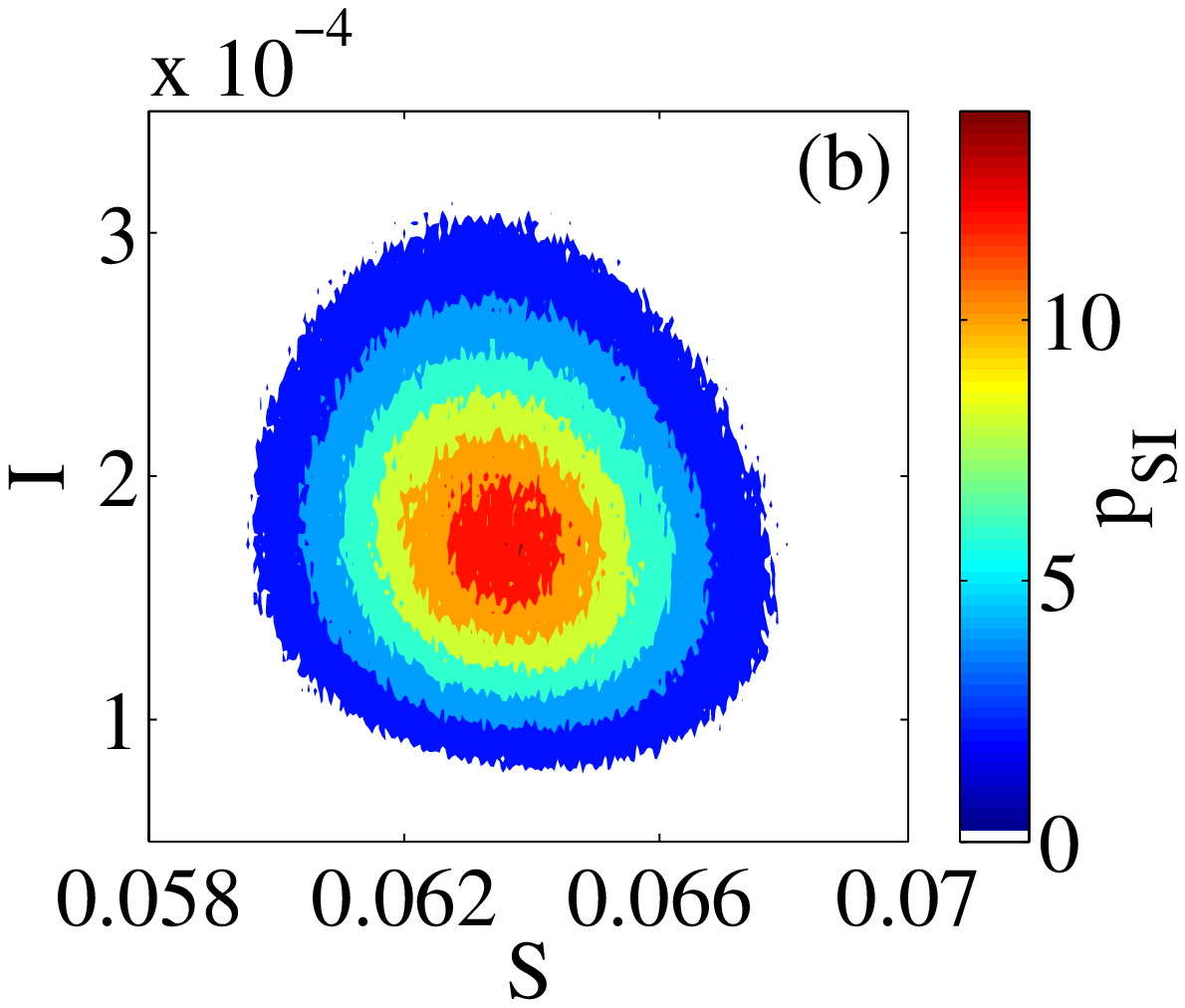}
\caption{\label{fig:simulationWSEI} (a) Time series of the fraction of the
  population that is infected with a disease, $I$, computed using a Markov
  simulation of the complete, transformed equations of the SEIR model
  [Eq.~(\ref{e:Vevolution})], and (b) (color online) a histogram of probability density,
  $p_{SI}$ of the $S$ and $I$ values found using a Markov
  simulation of Eq.~(\ref{e:Vevolution}).  The histogram is created using $100$
  years worth of data (starting with year $40$) for 500 realizations, and the
  color-bar values have been normalized by $10^{-4}$.}
\end{figure}

After performing a Markov simulation of Eq.~(\ref{e:Vevolution}) with a population size of $N=10$ million, we can compare the dynamics of the transformed system to the dynamics of the original system by transforming the $Y_i$ variables in the time series back to the
original $s$, $e$, and $i$ variables.  Dividing by $N$ yields $S$, $E$, and $I$.  Figure~\ref{fig:simulationWSEI}(a) shows a time series, after the transients have decayed, of the fraction
of the population that is infected with a disease, $I$.  The mean
and the frequency agree with those found from the Markov simulation of the
original system.  We have performed the simulation for
$500$ realizations, and a histogram representing the
probability density, $p_{SI}$ is
shown in Fig.~\ref{fig:simulationWSEI}(b).  The color-bar values in
Fig.~\ref{fig:simulationWSEI}(b) have been normalized by $10^{-4}$.  One can
see in Fig.~\ref{fig:simulationWSEI}(a) that the relative fluctuations of
the $I$ component has nearly doubled.  While the fluctuation size was 0.152 for the
original system, it is 0.310 for the transformed system.  Additionally, the
two histograms shown in Figs.~\ref{fig:simulationSEI}(b)
and~\ref{fig:simulationWSEI}(b) have a relative entropy of $d_{KL}=0.9519$,
which means they are not in agreement.  Because the simulation of the
stochastic dynamics in the complete system of transformed variables do not
qualitatively (or quantitatively) resemble the original stochastic system, we
cannot expect that the reduced system will agree with either the original or
the transformed systems.  Therefore, much care should be exercised when
extending the model reduction results (which show outstanding agreement)
derived for a specific type of noise in the limit of infinite population to
finite populations with a more general type of noise.

\section{Conclusions}\label{sec:conc}
We have considered the dynamics of an SEIR epidemiological model with stochastic forcing in the form
of additive, Gaussian noise.  We have presented two methods of model
reduction, whereby the goal is to project both the noise and the dynamics onto the
stochastic center manifold.  The first method uses the deterministic center
manifold found by neglecting the stochastic terms in the governing equations,
while the second method uses a stochastic normal form coordinate transform.

Since the original system of governing equations does not have
the necessary spectral structure to employ either deterministic or stochastic
center manifold theory, the system of equations has been
transformed using an appropriate linear transformation coupled with
appropriate parameter scaling.  At this stage, the first method of model
reduction can be performed by computing the deterministic center manifold
equation.  Substitution of this equation into the complete, stochastic system of transformed
equations leads to a reduced system of stochastic evolution equations.

The solutions of the complete, stochastic system of transformed equations as well as the reduced
system of equations were computed numerically.  We have shown that the
individual time series do not agree, because the noise has not been correctly projected onto
the stochastic center manifold.  However, by comparing histograms of the probability
density, $p_{SI}$ of the $S$ and $I$ values, we saw that there was very good agreement.
This is caused by the fact that although the two solutions are out of phase
with one another, their range of amplitude values are similar.  The phase difference is not
represented in the two histograms.  This is a real drawback when trying
to predict the timing of outbreaks, and leads to potential
problems when considering epidemic control, such as the enhancement of
disease extinction through random vaccine control~\cite{dyscla08}.

To accurately project the noise onto the manifold, we derived a stochastic normal form
coordinate transform for the complete, stochastic system of transformed
equations.  The numerical solution to this
reduced system was compared with the solution to the original system, and we
showed that there was excellent agreement, both qualitatively and
quantitatively.  As with the first method, the
histograms of the probability density, $p_{SI}$ of the $S$ and $I$ values agree very
well.

It should be noted that the use of these two reduction methods is not
constrained to problems in epidemiology, but rather may be used for many types
of physical problems.  For some generic systems, such as the singularly perturbed,
damped Duffing oscillator, either
reduction method can be used since the terms in the normal form coordinate
transform which lead to the average stochastic center manifold being different
from the deterministic center manifold occur at very high order~\cite{forsch09}.  In other
words, the average stochastic center manifold and deterministic center
manifold are virtually identical.  For the SEIR model considered in this
article, there are terms at low order in the normal form transform which cause
a significant difference between the average stochastic center manifold and
the deterministic manifold.  Therefore, as we have demonstrated, when working
with the SEIR model, one must use
the normal form coordinate transform method to correctly project the noise
onto the center manifold.    

In summary, we have presented a new method of stochastic model
reduction that allows for impressive
improvement in time series prediction.  The reduced model captures both the
amplitude and phase accurately for a temporal scale that is many orders of
magnitude longer than the typical relaxation time.  Since sufficient statistics
of disease data are limited due to short time series collection, the results
presented here
provide a potential  method to properly model real, stochastic disease data in
the time domain.  Such  long-term accuracy of the reduced model will allow for the
application of effective control of a disease where phase differences between
outbreak times and vaccine controls are important.  Additionally, since
our method is general, it may be
applied to very high-dimensional epidemic models, such as those involving
adaptive networks.  From a dynamical systems viewpoint, the reduction  method has the
potential to  accurately capture new, emergent dynamics as we increase the size of the random fluctuations. This could be a
means to identify  new noise-induced phenomena in generic stochastic
systems.

\section*{Acknowledgments}
The authors benefited from the comments and suggestions of anonymous
reviewers.  We gratefully acknowledge support from the Office of Naval
Research and the Air Force Office of Scientific Research.  E.F. is supported by a
National Research Council Research Fellowship, and L.B. is supported by award number R01GM090204 from the National Institute Of
General Medical Sciences. The content is solely the responsibility of the
authors and does not necessarily represent the official views of the National
Institute Of General Medical Sciences or the
National Institutes of Health.

\begin{appendix}
\section{Details of the Iterative Procedure for a Simple Example}\label{sec:simple_example}
We consider the system given by 
\begin{subequations}
\begin{flalign}
\frac{dx}{d\tau}=&\mu(y+\sigma\phi),\label{e:Duff_trans_x_stoch}\\
\frac{dy}{d\tau}=&x-x^3-y,\label{e:Duff_trans_y_stoch}\\
\frac{d\mu}{d\tau}=&0.\label{e:Duff_trans_eps_stoch}
\end{flalign}
\end{subequations}  

The iterative procedure begins by letting
\begin{subequations}
\begin{equation}
\label{e:xNF}
x\approx X,
\end{equation}
\begin{equation}
\label{e:XdotNF}
X^{\prime}\approx 0,
\end{equation}
\end{subequations}
and by finding a change to the $y$ coordinate (fast process) with the form
\begin{subequations}
\begin{equation}
\label{e:yNF}
y = Y+\eta(\tau,X,Y)+\ldots,
\end{equation}
\begin{equation}
\label{e:YdotNF}
Y^{\prime} = -Y+G(\tau,X,Y)+\ldots,
\end{equation}
\end{subequations}
where $\eta$ and $G$ are small corrections to the coordinate transform and the
corresponding evolution equation.  Substitution of
Eqs.~(\ref{e:xNF})-(\ref{e:YdotNF}) into Eq.~(\ref{e:Duff_trans_y_stoch})
gives the equation
\begin{equation}
Y^{\prime}+\frac{\partial \eta}{\partial \tau} + \frac{\partial \eta}{\partial
  X}\frac{\partial X}{\partial \tau} + \frac{\partial \eta}{\partial
  Y}\frac{\partial Y}{\partial \tau} = -Y -\eta +X-X^3.
\end{equation}
Replacing $Y^{\prime}=\partial Y/\partial \tau$ with $-Y+G$ [Eq.~(\ref{e:YdotNF})],
noting that $\partial X/\partial \tau =0$ [Eq. (\ref{e:XdotNF})], and ignoring the
term $\partial\eta/\partial Y\cdot G$ since it is a product of small
corrections leads to 
\begin{equation}
\label{e:Getaeqn}
G+\frac{\partial \eta}{\partial \tau} -Y \frac{\partial \eta}{\partial
  Y}+\eta = X-X^3.
\end{equation}

Equation~(\ref{e:Getaeqn}) must now be solved for $G$ and $\eta$.  In order to
keep the evolution equation [Eq.~(\ref{e:YdotNF})] as simple as possible
(principle~(\ref{four}) of Sec.~\ref{sec:correct_proj}), we let
$G=0$, which means that the coordinate transform [Eq.~(\ref{e:yNF})] is modified by $\eta =X-X^3$.
Therefore, the new approximation of the coordinate transform and its dynamics
are given by
\begin{subequations}
\begin{equation}
\label{e:yNF2_a}
y = Y+X-X^3+\mathcal{O}(\zeta^2),
\end{equation}
\begin{equation}
\label{e:yNF2_b}
Y^{\prime} = -Y+\mathcal{O}(\zeta^2),
\end{equation}
\end{subequations}
where $\zeta =|(X,Y,\mu,\sigma)|$ so that $\zeta$ provides a count of the
number of $X$, $Y$, $\mu$, and $\sigma$ factors in any one term. 

For the second iteration, we seek a correction to the $x$ coordinate (slow process)
with the form 
\begin{subequations}
\begin{equation}
\label{e:xNF2}
x = X+\xi(\tau,X,Y)+\ldots, 
\end{equation}
\begin{equation}
\label{e:XdotNF2}
X^{\prime} = F(\tau,X,Y)+\ldots,
\end{equation}
\end{subequations}
where $\xi$ and $F$ are small corrections.  Substitution of
Eqs.~(\ref{e:yNF2_a})-(\ref{e:XdotNF2}) into
Eq. (\ref{e:Duff_trans_x_stoch}) leads to
\begin{equation}
X^{\prime}+\frac{\partial \xi}{\partial \tau} + \frac{\partial \xi}{\partial
  X}\frac{\partial X}{\partial \tau} + \frac{\partial \xi}{\partial
  Y}\frac{\partial Y}{\partial \tau} = \mu(Y+X-X^3)+\mu\sigma\phi.
\end{equation}
Replacing $X^{\prime}=\partial X/\partial \tau$ with $F$ [Eq.~(\ref{e:XdotNF2})],
replacing $\partial Y/\partial \tau$ with $-Y$ [Eq.~(\ref{e:yNF2_b})], and
ignoring the term $\partial\xi/\partial X\cdot F$ since it is a product of
small corrections gives the equation:
\begin{equation}
\label{e:Fxieqn}
F+\frac{\partial \xi}{\partial \tau} -Y\frac{\partial \xi}{\partial
  Y} = \mu(Y+X-X^3)+\mu\sigma\phi.
\end{equation}

Equation~(\ref{e:Fxieqn}) must now be solved for $F$ and $\xi$.  As in the
first step, we employ principle~(\ref{four}) and keep the
evolution equation [Eq.~(\ref{e:XdotNF2})] as simple as possible.  However, since the terms
$\mu (X-X^3)$ located on the right-hand side of Eq.~(\ref{e:Fxieqn}) do not contain $\tau$ or
$Y$, these terms must be included in F.  Therefore, one piece of $F$ will be
$F=\mu (X-X^3)$.  

The remaining deterministic term on the right-hand side
of Eq.~(\ref{e:Fxieqn}) contains $Y$.  This term can therefore be integrated
into $\xi$.  The equation to be solved is 
\begin{equation}
\label{e:Eq_2_a}
-Y \frac{\partial\xi}{\partial Y}=\mu Y, 
\end{equation}
whose solution is given as $\xi =-\mu Y$.

To abide by principle~(\ref{four}), we would like to integrate the
stochastic piece on the right-hand side of Eq.~(\ref{e:Fxieqn}) into $\xi$, by
solving the equation
\begin{equation}
\label{e:Eq_2_b}
\partial \xi/\partial \tau = \mu\sigma\phi.  
\end{equation}
However, the solution of Eq.~(\ref{e:Eq_2_b}) is given by
\begin{equation}
\xi=\mu\sigma\int\phi \, d\tau,
\end{equation}
which has secular growth like a Wiener
process.  Since this would violate principle~(\ref{one}), 
we must let
$F=\mu\sigma\phi$.

Putting the three pieces together yields $\xi=-\mu Y$ and
$F=\mu(X-X^3)+\mu\sigma\phi$.  Therefore, the new approximation of the coordinate transform and its dynamics
are given by
\begin{subequations}
\begin{equation}
\label{e:xNF3_a}
x=X-\mu Y+\mathcal{O}(\zeta^3),
\end{equation}
\begin{equation}
\label{e:xNF3_b}
X^{\prime}=\mu(X-X^3)+\mu\sigma\phi+\mathcal{O}(\zeta^3).
\end{equation}
\end{subequations}

The construction of the normal form continues by seeking corrections, $\xi$ and $F$, to the
$x$ coordinate transform and the $X$ evolution using the
updated residual of the $x$ equation [Eq.~(\ref{e:Duff_trans_x_stoch})], and
by seeking corrections, $\eta$ and $G$, to the
$y$ coordinate transform and the $Y$ evolution equation using the
updated residual of the $y$ equation [Eq.~(\ref{e:Duff_trans_y_stoch})].

\section{Reduced, stochastic SEIR model:  Correct projection of the noise
}\label{sec:cpcm_mort}
The specific form of $F$ and $G$ in Eqs.~(\ref{e:F}) and~(\ref{e:G}) are given as 
\begin{subequations}
\begin{flalign}
F =& -
\left [\alpha_0^2\gamma_0^3X_2  + \frac{\mu \beta \alpha_0^2}{\left(\alpha_0+\gamma_0\right )} \gamma_0^2 \left (-\frac{\gamma_0^2 }{\alpha_0+\gamma_0} X_2^2 + \alpha_0 X_1X_2\right )+ \mu^2\left (\alpha_0\gamma_0^3X_1 +
\right. \right.&& \nonumber\\
& \left. \frac{\alpha_0\gamma_0^2\left ( 2\alpha_0^3+5\alpha_0^2\gamma_0+5\alpha_0\gamma_0^2+\gamma_0^3\right )}{\left (\alpha_0+\gamma_0\right )^2}X_2 - \frac{\alpha_0^2\beta^2\gamma_0^2}{\left ( \alpha_0+\gamma_0 \right
  )}X_1^2X_2-  \frac{\alpha_0^2\beta^2\gamma_0^3}{\left (\alpha_0+\gamma_0\right
  )^2}X_1X_2^2 \right )+\nonumber\\
&\mu^3\left (\alpha_0^2\beta\gamma_0X_1
 -\frac{\alpha_0^2\beta\gamma_0^3}{\left (\alpha_0+\gamma_0\right )^2}X_2
 +\frac{\alpha_0^2\beta\gamma_0^2}{\left (\alpha_0+\gamma_0\right )}X_1^2
 -\frac{3\alpha_0\beta\gamma_0^3}{\left (\alpha_0+\gamma_0\right
  )}X_2^2 +\right.\nonumber\\
& \left.\left. \frac{\alpha_0\beta\gamma_0\left (
   \alpha_0^3-\alpha_0^2\gamma_0 -3\alpha_0\gamma_0^2-3\gamma_0^3\right
  )}{\left (\alpha_0+\gamma_0\right )^2}X_1X_2 \right )\right ]
/ \left[\gamma_0\left (\alpha_0+\gamma_0\right )\left
   (\alpha_0+\mu^2\right )\left (\gamma_0+\mu^2\right )\right] 
+ \nonumber \\
& \sigma_5 \phi_5 -  \frac{\mu^2 (\alpha_0^2 +\alpha_0\gamma_0+\gamma_0^2)}{(\alpha_0+\gamma_0)^3}
\left( \frac{\sigma_4\alpha_0\phi_4}{\gamma_0} + \frac{\sigma_6\gamma_0\phi_6}{\alpha_0+\gamma_0} \right ) - \nonumber \\
& \frac{\mu^3\alpha_0\beta}{(\alpha_0+\gamma_0)^3}
\left( \frac{\sigma_4\alpha_0\phi_4}{\gamma_0} +\frac{\sigma_6\gamma_0\phi_6}{(\alpha_0+\gamma_0)} \right)
,\label{e:x1_NFCT_mort}
\end{flalign}
\begin{flalign}
G=& \left [\mu \left (\frac{\alpha_0^3\beta\gamma_0^2}{\left
    (\alpha_0+\gamma_0\right )}X_1X_2
  -\frac{\alpha_0^2\beta\gamma_0^4}{\left (\alpha_0+\gamma_0\right
   )^2}X_2^2 \right ) +  \mu^2\left ( -\alpha_0^2\gamma_0^2X_1 +\frac{\alpha_0^2\gamma_0^4}{\left(\alpha_0+\gamma_0\right)^2}X_2-   \right. \right.  &&\nonumber\\
& \left. \frac{\alpha_0^2\beta^2\gamma_0^2}{\left (\alpha_0+\gamma_0\right
  )}X_1^2X_2-\right.\left.\frac{\alpha_0^2\beta^2\gamma_0^3}{\left
   (\alpha_0+\gamma_0\right )^2}X_1X_2^2\right )+ \mu^3\left ( \alpha_0^2\beta\gamma_0 X_1-\frac{\alpha_0^2\beta\gamma_0^3}{\left (\alpha_0+\gamma_0\right )^2}X_2
 + \right.  \nonumber\\
&  \left. \frac{\alpha_0^2\beta\gamma_0^2\left (\alpha_0+\mu^2\right )\left
   (\gamma_0+\mu^2\right )}{\left (\alpha_0+\gamma_0\right )}X_1^2
 - \frac{3\alpha_0\beta\gamma_0^3}{\left
   (\alpha_0+\gamma_0\right )}X_2^2 -3\alpha_0\beta\gamma_0^2\left
  (\alpha_0+\mu^2\right )\left (\gamma_0+\mu^2\right )X_1X_2+ \right.\nonumber\\
&   \left.\left. \frac{\alpha_0^2\beta\gamma_0\left
  (\alpha_0^2+2\alpha_0\gamma_0+3\gamma_0^2\right )}{\left
  (\alpha_0+\gamma_0\right )^2}X_1X_2\right )\right ]/\left
 [\alpha_0\gamma_0\left
   (\alpha_0+\mu^2\right )\left (\gamma_0+\mu^2\right )\right ]+ \sigma_6\phi_6 +
\nonumber\\
& \frac{\mu^2\sigma_4\left
  (\alpha_0^2+\alpha_0\gamma_0+\gamma_0^2 \right )}{\alpha_0\gamma_0\left
  (\alpha_0+\gamma_0\right )}\phi_4 +\frac{\mu^2\sigma_6\left [\gamma_0^3\left (\alpha_0+\mu^2\right )\left
  (\gamma_0+\mu^2\right ) +\alpha_0\gamma_0\left (\alpha_0+\gamma_0\right
 )\right ]}{\alpha_0\left (\alpha_0+\gamma_0\right )^3}\phi_6+\nonumber\\
&\frac{\mu^3\beta}{\left (\alpha_0+\gamma_0 \right )}\left
 (\frac{\sigma_4\phi_4}{\gamma_0} +\frac{\sigma_6\gamma_0\phi_6}{\left
   (\alpha_0+\gamma_0\right )^2}\right ).\label{e:x2_NFCT_mort}
\end{flalign}
\end{subequations}

\section{Markov simulation for transformed SEIR model}\label{sec:MSim}

The complete system in the transformed variables has the stable endemic
equilibrium at the origin. To bound the dynamics to the first octant, we
transform the new variables by using the original properties of $s \ge 0$, $e
\ge 0$, and $i \ge 0$, so that 
\begin{equation}
\label{e:ineq}
u \le {\frac {\mu^2  N\gamma_0}{\alpha_0  \left( \alpha_0+\gamma_0 \right) }}, ~~~ 
 -{\frac {N\gamma_0  \left( \beta {\mu}^{3}+{\alpha_0}^{2}+\gamma_0 \alpha_0 \right) }{\alpha_0   \beta \mu \left( \alpha_0 +\gamma_0 \right)}} \le v, 
~~~ -{\frac {N \mu^2  \left( \alpha_0+\gamma_0 \right) }{\gamma_0 \alpha_0}}
\le w. 
\end{equation}
Therefore, we define the following new variables:
\begin{subequations}
\begin{eqnarray}
Y_1 &=& -u+{\frac {N \mu^2 \gamma_0}{\alpha_0  \left( \alpha_0+\gamma_0 \right) }}, \label{e:Y1}\\
Y_2 &=& v+ {\frac {N\gamma_0  \left( \beta {\mu}^{3}+{\alpha_0}^{2}+\gamma_0 \alpha_0 \right) }{\alpha_0   \beta \mu \left( \alpha_0 +\gamma_0 \right)}},\label{e:Y2}\\
Y_3 &=& w+{\frac {N  \mu^2  \left( {\it \alpha_0}+\gamma_0 \right) }{\gamma_0 \alpha_0}}.\label{e:Y3}
\end{eqnarray}
\end{subequations}
In these variables, we define evolution relationships similar to Eq.~(\ref{e:evolution}). The complete transformed system will evolve in $\tau$ the following way: 
\begin{equation} \label{e:Vevolution}
\begin{array}{lcl}
\mbox{transition}    &\hspace*{0.25in} &\mbox{rate}  \\
(Y_1+1,Y_2,Y_3)  && \frac{\beta \mu}{N}\left(\frac{\gamma_0}{\alpha_0+\gamma_0}Y_2 Y_3+ Y_1^2\right)\\
(Y_1-1,Y_2,Y_3)  && (\alpha_0+\mu^2)Y_1+\frac{\beta \mu}{N}\left(\frac {\gamma_0 }{\alpha_0+\gamma_0}Y_1 Y_3   +Y_1 Y_2 \right) \\
(Y_1,Y_2+1,Y_3)  && \mu^2 N 
+\frac{\beta \mu}{N}\left(
{\frac {\alpha_0  }{ \left( \alpha_0+\gamma_0 \right) } Y_1 Y_3}
+{\frac {\alpha_0  }{\gamma_0}}Y_1 Y_2  \right) \\(Y_1,Y_2-1,Y_3) && \alpha_0 Y_1+\mu^2 Y_2
+\frac{\beta \mu}{N}\left( {\frac {\alpha_0  }{ \left( \alpha_0+\gamma_0 \right) }}  Y_2 Y_3 +{\frac {\alpha_0 }{\gamma_0}} {Y_1}^{2} \right) \\
(Y_1,Y_2,Y_3+1) && (\alpha_0+\gamma_0) Y_1
+ \frac{\beta \mu}{N}\left(   Y_2 Y_3+{\frac {\left( \alpha_0+\gamma_0 \right)}{\gamma_0}}  {Y_1}^{2} \right)   \\
(Y_1,Y_2,Y_3-1) && (\gamma_0+\mu^2) Y_3 +\frac{\beta \mu}{N}\left(  Y_1 Y_3 +{\frac { \left( \alpha_0+\gamma_0 \right) }{\gamma_0}} Y_1 Y_2 \right)
\end{array} .
\end{equation} 

\end{appendix}


\end{document}